\documentclass[11pt]{article}

\usepackage{calc}
\usepackage{rotating}
\usepackage[english]{babel}
\usepackage{graphicx}
\usepackage{subfig}
\usepackage{float}
\usepackage{amsmath}
\usepackage{amssymb}
\usepackage{amsthm}
\usepackage{latexsym}
\usepackage{dcolumn}
\usepackage{multirow}
\usepackage{geometry}
\usepackage{color}
\usepackage{hyperref}
\usepackage{mciteplus}
\usepackage{soul}

\def\gtwid{\mathrel{\raise.3ex\hbox{$>$\kern-.75em\lower1ex\hbox{$\sim
$}}}}
\def\vio{\mathrel{\hbox{$E$\kern-.60em\hbox{$/
$}}}}
\newcommand{\newc}{\newcommand*}

\long\def\begincomment#1\endcomment{%
        \begingroup\sf\baselineskip12pt#1\endgroup}

\newc{\hobs}{\ensuremath{H_\text{obs}}}

\newc{\etal}{\textrm{et al.}} 
\newc{\eg}{\textrm{e.g.}} 
\newc{\ie}{\textrm{i.e.}}
\newc{\etc}{\textrm{etc}}
\newc\vs{\textrm{vs.}}
\newc{\cl}{\rm {CL}}

\newc{\ev}{\ensuremath{\,\mathrm{eV}}}
\newc{\kev}{\ensuremath{\,\mathrm{keV}}}
\newc{\mev}{\ensuremath{\,\mathrm{MeV}}}
\newc{\gev}{\ensuremath{\,\mathrm{GeV}}}
\newc{\tev}{\ensuremath{\,\mathrm{TeV}}}

\newc{\MeV}{\mev} 
\newc{\TeV}{\tev}
\newc{\invpb}{\ensuremath{/\text{pb}}}
\newc{\invfb}{\ensuremath{/\text{fb}}}

\newc\nb{\ensuremath{\,\mathrm{nb}}} \newc\pb{\ensuremath{\,\mathrm{pb}}} \newc\fb{\ensuremath{\,\mathrm{fb}}}

\newc\pc{\ensuremath{\,\mathrm{pc}}}
\newc\kpc{\ensuremath{\,\mathrm{kpc}}}
\newc\mpc{\ensuremath{\,\mathrm{Mpc}}}

\newc\ps{\ensuremath{\,\mathrm{ps}}} 

\newc\cmeter{\ensuremath{\,\mathrm{cm}}} 
\newc\meter{\ensuremath{\,\mathrm{m}}} 
\newc\kmeter{\ensuremath{\,\mathrm{km}}}

\newc\second{\ensuremath{\,\mathrm{s}}}
\newc\msecond{\ensuremath{\,\mathrm{ms}}}
\newc\nsecond{\ensuremath{\,\mathrm{ns}}}
\newc\psecond{\ensuremath{\,\mathrm{ps}}}

\newc{\chisqmin}{\ensuremath{\chi^2_{\mathrm{min}}}}
\newc{\Delchisq}{\ensuremath{\Delta\chi^2}}
\newc{\chisq}{\ensuremath{\chi^2}}

\newc{\like}{\ensuremath{\mathcal{L}}}

\newc\lsim{\ensuremath{\mathrel{\rlap{\lower4pt\hbox{\hskip1pt$\sim$}}\raise1pt\hbox{$<$}}}}
\newc\gsim{\ensuremath{\mathrel{\rlap{\lower4pt\hbox{\hskip1pt$\sim$}}\raise1pt\hbox{$>$}}}}

\newc{\VEV}[1]{\ensuremath{\langle #1 \rangle}}

\newc{\dl}{\ensuremath{\stackrel{\leftarrow}{D}}}
\newc{\dr}{\ensuremath{\stackrel{\rightarrow}{D}}}

\newc{\bcenter}{\begin{center}}    \newc{\ecenter}{\end{center}}
\newc{\bfl}{\begin{flushleft}}    \newc{\efl}{\end{flushleft}}
\newc{\bfr}{\begin{flushright}}    \newc{\efr}{\end{flushright}}

\newc{\bi}{\begin{itemize}}
\newc{\ei}{\end{itemize}}
\newc{\bed}{\begin{description}}
\newc{\eed}{\end{description}}
\newc{\ben}{\begin{enumerate}}
\newc{\een}{\end{enumerate}}

\newc{\be}{\begin{equation}}
\newc{\ee}{\end{equation}}
\newc{\bea}{\begin{eqnarray}}
\newc{\eea}{\end{eqnarray}}
\newc{\ra}{\rightarrow}

\newc{\alphas}{\ensuremath{\alpha_s}}
\newc{\alphatwo}{\ensuremath{\alpha_2}}
\newc{\alphaone}{\ensuremath{\alpha_1}}
\newc{\alphai}[1]{\ensuremath{\alpha_{#1}}}
\newc{\alphaem}{\ensuremath{\alpha_{\mathrm{em}}}}

\newc{\alphaeff}{\ensuremath{\alpha_{\mathrm{eff}}}}

\newc{\sineff}{\ensuremath{\sin \theta_{\mathrm{eff}}}}
\newc{\sinsqeff}{\ensuremath{\sin^2 \theta_{\mathrm{eff}}}}
\newc{\dalphahad}{\ensuremath{\Delta \alpha_{\mathrm{had}}}}

\newc{\yt}{\ensuremath{h_t}} \newc{\yb}{\ensuremath{h_b}} \newc{\ytau}{\ensuremath{h_{\tau}}}

\newc\mz{\ensuremath{M_Z}} 
\newc\mw{\ensuremath{m_W}}
\newc\mZ{\mz}        \newc\mW{\mw}

\newc\mhsm{\ensuremath{ m_{H_{\mathrm{SM}}}}}

\newc{\mtop}{\ensuremath{ m_t}}               \newc{\mtpole}{\ensuremath{ M_t}}
\newc{\mbottom}{\ensuremath{ m_b}} 
\newc{\mtau}{\ensuremath{ m_{\tau}}}
\newc{\mt}{\mtpole}
\newc{\mb}{\mbottom} 

\newc{\rtwogg}{\ensuremath{R_{h_2}(\gamma\gamma)}}
\newc{\rtwozz}{\ensuremath{R_{h_2}(ZZ)}}
\newc{\ronegg}{\ensuremath{R_{h_1}(\gamma\gamma)}}
\newc{\ronezz}{\ensuremath{R_{h_1}(ZZ)}}
\newc{\rsiggg}{\ensuremath{R_{h_\textrm{sig}}(\gamma\gamma)}}
\newc{\rsigzz}{\ensuremath{R_{h_\textrm{sig}}(ZZ)}}

\newc{\llbar}{\ensuremath{\ell\bar{\ell}}}
\newc{\tauptaum}{\ensuremath{ \tau^+\tau^-}}

\newc{\qqbar}{\ensuremath{ q\bar{q}}} \newc{\ppbar}{\ensuremath{ p\bar{p}}}
\newc{\bbbar}{\ensuremath{ b\bar{b}}} \newc{\ttbar}{\ensuremath{ t\bar{t}}}
\newc{\ffbar}{\ensuremath{ f\bar{f}}} \newc{\tautaubar}{\ensuremath{ \tau\bar{\tau}}}

\newc{\mchi}{\ensuremath{m_\neutone}}
\newc{\squark}{\ensuremath{\tilde{q}}}
\newc{\slepton}{\ensuremath{\tilde{l}}}
\newc{\gluino}{\ensuremath{\tilde{g}}} 
\newc{\mgluino}{\ensuremath{{m_{\gluino}}}}

\newc{\sthw}{\ensuremath{ \sin\theta_W}}              \newc{\cthw}{\ensuremath{\cos\theta_W}}
\newc{\tanthw}{\ensuremath{ \tan\theta_W}}              \newc{\cotthw}{\ensuremath{\cot\theta_W}}

\newc{\ssqthw}{\ensuremath{\sin^2 \theta_W}}

\newc{\msbar}{\ensuremath{\overline{MS}}} \newc{\drbar}{\ensuremath{\overline{DR}}}

\newc{\mtmtsmmsbar}{\ensuremath{ m_t(m_t)^{\msbar}_{{\mathrm{SM}}}}}
\newc{\mtmtsmdrbar}{\ensuremath{ m_t(m_t)^{\drbar}_{{\mathrm{SM}}}}}
\newc{\mtmtmssmdrbar}{\ensuremath{ m_t(m_t)^{\drbar}_{{\mathrm{SUSY}}}}}

\newc{\mbmbmsbar}{\ensuremath{ m_b(m_b)^{\msbar} }}

\newc{\mbmbsmmsbar}{\ensuremath{ m_b(m_b)^{\msbar}_{{\mathrm{SM}}}}}
\newc{\mbmzsmmsbar}{\ensuremath{ m_b(\mz)^{\msbar}_{{\mathrm{SM}}}}}
\newc{\mbmzsmdrbar}{\ensuremath{ m_b(\mz)^{\drbar}_{{\mathrm{SM}}}}}
\newc{\mbmzmssmdrbar}{\ensuremath{ m_b(\mz)^{\drbar}_{{\mathrm{SUSY}}}}}

\newc{\mtaumzsmmsbar}{\ensuremath{ m_{\tau}(\mz)^{\msbar}_{{\mathrm{SM}}}}}
\newc{\mtaumzsmdrbar}{\ensuremath{ m_{\tau}(\mz)^{\drbar}_{{\mathrm{SM}}}}}
\newc{\mtaumzmssmdrbar}{\ensuremath{ m_{\tau}(\mz)^{\drbar}_{{\mathrm{SUSY}}}}}

\newc{\alphasmzms}{\ensuremath{\alpha_s(M_Z)^{\overline{MS}}}}
\newc{\alphaimzms}[1]{\ensuremath{\alpha_{#1}(M_Z)^{\overline{MS}}}}

\newc{\alphaemmz}{\ensuremath{\alpha_{\mathrm{em}}(M_Z)^{\overline{MS}}}}

\newc{\mzero}{\ensuremath{{m_0}}}
\newc{\mhalf}{\ensuremath{ m_{1/2}}}
\newc{\tanb}{\ensuremath{\tan\beta}}
\newc{\azero}{\ensuremath{ A_0}}
\newc{\signmu}{\ensuremath{\rm{sgn}\,\mu}}
\newc{\atau}{\ensuremath{{A_{\tau}}}}

\newc{\mueff}{\ensuremath{\mu_{\rm{eff}}}}

\newc{\lam}{\ensuremath{{\lambda}}}
\newc{\kap}{\ensuremath{{\kappa}}}

\newc{\alam}{\ensuremath{{A_{\lambda}}}}
\newc{\akap}{\ensuremath{{A_{\kappa}}}}

 \newc{\hsm}{\ensuremath{H_{\rm SM}}}      

\newc{\mgut}{\ensuremath{ M_{\rm GUT}}}
\newc{\mplanck}{\ensuremath{ M_{\rm P}}}      \newc{\mpl}{\ensuremath{ M_{\rm Pl}}}
\newc{\msusy}{\ensuremath{ M_{\rm SUSY}}}      \newc{\ms}{\ensuremath{ M_{\rm S}}}

 \newc{\hu}{\ensuremath{ H_u}}       \newc{\hd}{\ensuremath{ H_d}}
 \newc{\mhu}{\ensuremath{ m_{H_u}}}       \newc{\mhd}{\ensuremath{ m_{H_d}}}
 \newc{\mhuew}{\ensuremath{ m^{\ast}_{H_u}}}       \newc{\mhdew}{\ensuremath{ m^{\ast}_{H_d}}}
 \newc{\mhuewsq}{\ensuremath{ m^{\ast\, 2}_{H_u}}}       \newc{\mhdewsq}{\ensuremath{ m^{\ast\, 2}_{H_d}}}
 \newc{\mhl}{\ensuremath{m_\hl}} 
 \newc{\mhone}{\ensuremath{m_{h_1}}} 
 \newc{\mhtwo}{\ensuremath{m_{h_2}}} 
 \newc{\mglu}{\ensuremath{m_{\tilde g}}} 
 \newc{\mul}{\ensuremath{m_{\tilde{u}_L}}} 
 \newc{\mtone}{\ensuremath{m_{\tilde{t}_1}}} 
 \newc{\ma}{\ensuremath{m_A}} 
 \newc{\maone}{\ensuremath{m_{a_1}}} 
 \newc{\matwo}{\ensuremath{m_{a_2}}}
 \newc{\hone}{\ensuremath{h_1}}
 \newc{\htwo}{\ensuremath{h_2}}
 \newc{\aone}{\ensuremath{a_1}}
 \newc{\atwo}{\ensuremath{a_2}}

\newc{\sigsip}{\ensuremath{\sigma^{\rm SI}_{p}}}	\newc{\sigsin}{\ensuremath{\sigma^{\rm SI}_{n}}}
\newc{\sigsdp}{\ensuremath{\sigma^{\rm SD}_{p}}}	\newc{\sigsdn}{\ensuremath{\sigma^{\rm SD}_{n}}}
\newc{\sigsi}{\ensuremath{\sigma^{\rm SI}}}	\newc{\sigsd}{\ensuremath{\sigma^{\rm SD}}}

\newc{\abund}{\ensuremath{ \Omega h^2}}
\newc{\omegadm}{\ensuremath{ \Omega_{{\rm DM}}}}     \newc{\abunddm}{\ensuremath{ \Omega_{{\rm DM}} h^2}} 
\newc{\omegam}{\ensuremath{ \Omega_{{\rm m}}}}       \newc{\abundm}{\ensuremath{ \Omega_{{\rm m}} h^2}}
\newc{\omegab}{\ensuremath{ \Omega_{{\rm b}}}}	\newc{\abundb}{\ensuremath{ \Omega_{{\rm b}} h^2}}
\newc{\omegatot}{\ensuremath{ \Omega_{{\rm TOT}}}}
\newc{\omegacdm}{\ensuremath{ \Omega_{{\rm CDM}}}}   \newc{\abundcdm}{\ensuremath{ \Omega_{{\rm CDM}} h^2}}
\newc{\omegalambda}{\ensuremath{ \Omega_{\Lambda}}} \newc{\abundlambda}{\ensuremath{ \Omega_{\Lambda} h^2}}
\newc{\omegarad}{\ensuremath{ \Omega_{{\rm rad}}}}  \newc{\abundrad}{\ensuremath{ \Omega_{{\rm rad}} h^2}}

\newc{\rhocrit}{\ensuremath{ \rho_{\rm crit}}}
\newc{\rhochi}{\ensuremath{ \rho_{\chi}}}

\newc{\abunchi}{\ensuremath{\Omega_\chi h^2}}
\newc{\abundlsp}{\ensuremath{\Omega_{\rm LSP}h^2}}

\newc{\amu}{\ensuremath{ a_{\mu}}}        \newc{\amususy}{\ensuremath{ a_{\mu}^{\mathrm{SUSY}}}}
\newc{\amuexpt}{\ensuremath{ a_{\mu}^{\mathrm{expt}}}}        \newc{\amusm}{\ensuremath{ a_{\mu}^{\mathrm{SM}}}}
\newc\deltaamu{\ensuremath{\Delta a_{\mu}}} \newc{\deltaamususy}{\ensuremath{\delta a_{\mu}^{\mathrm{SUSY}}}}
\newc\gmtwo{\ensuremath{ (g-2)_{\mu}}} 
\newc{\deltagmtwomususy}{\ensuremath{\delta\left(g-2\right)_{\mu}^{\mathrm{SUSY}}}}
\newc{\deltagmtwomu}{\ensuremath{\delta\left(g-2\right)_{\mu}}}

\newc\BR{\ensuremath{\rm BR}}

\newc\bsgamma{\ensuremath{ b\rightarrow s \gamma }}
\newc\bxsgamma{\ensuremath{\overline{B}\rightarrow X_{s}\gamma}}

\newc\brbsgamma{\ensuremath{\BR\left(\bsgamma\right)}}
\newc\brbxsgamma{\ensuremath{\BR\left(\bxsgamma\right)}}

\newc\bsmumu{\ensuremath{B_s\to\mu^+\mu^-}}
\newc\brbsmumu{\ensuremath{\BR\left(B_s\to\mu^+\mu^-\right)}}

\newc\bdmmumu{\ensuremath{\overline{B}_d\to\mu^+\mu^-}}

\newc\bbbarmix{\ensuremath{\overline{B}_s\mbox{-}B_s}}      
\newc\delmbs{\ensuremath{\Delta M_{B_s}}}

\newc{\butaunu}{\ensuremath{B_u \rightarrow \tau \nu}}
\newc{\brbutaunu}{\ensuremath{\BR\left(B_u \rightarrow \tau \nu\right)}}

     \newcommand*{\refsec}[1]{Sec.~\ref{#1}}

\newcommand*{\neutone}{\ensuremath{\chi}}

\newcommand*{\micromegas}{MicrOMEGAs}

\newcommand*{\superiso}{\text{SuperIso}}

\newcommand*{\nmssmtools}{\text{NMSSMTools}}
\newcommand{\half}{\tfrac{1}{2}}
\let\oldcite\cite
\renewcommand*{\cite}{~\oldcite}

\renewcommand*{\hl}{\ensuremath{h}}

\newcommand*{\hpm}{\ensuremath{H^{\pm}}}
\newcommand*{\mhpm}{\ensuremath{m_{\hpm}}}

\begin{document}

\thispagestyle{empty}
\vspace*{2.0cm}
\begin{center}
{\Large \bf {Charged Higgs boson in the $\mathbf{W^\pm}$~Higgs channel \\[0.25cm]
at the Large Hadron Collider}}\\ 
\vspace{.3in}
{\large Rikard Enberg$^a$, William Klemm$^a$, Stefano
  Moretti$^b$, Shoaib Munir$^{a,c}$ and Glenn Wouda$^a$} \\[0.25cm]
{\sl $^a$ Department of Physics and Astronomy, \\ 
Uppsala University, Box 516, SE-751 20 Uppsala, Sweden.}\\[0.25cm]
{\sl $^{b}$ School of Physics \& Astronomy, \\
University of Southampton, Southampton SO17 1BJ, UK.} \\[0.25cm]
{\sl $^{c}$ Asia Pacific Center for Theoretical Physics, San 31, Hyoja-dong, \\
  Nam-gu, Pohang 790-784, Republic of Korea.} \\[0.25cm]\end{center}
\vspace{0.3in}

\begin{abstract} 
In light of the recent discovery of a neutral Higgs boson, \hobs, with
a mass near 125\gev, we reassess the LHC discovery potential of
a charged Higgs boson, \hpm, in the $W^\pm \hobs$
decay channel. This decay channel can be
particularly important for a \hpm\ heavier than the top quark,
when it is produced through the $pp\to t\hpm$ process. 
The knowledge of the mass of \hobs\
 provides an additional handle in the kinematic selection when 
reconstructing a Breit-Wigner resonance in the
$\hobs\to b\bar b$ decay channel.
 We consider some  extensions of the
Standard Model Higgs sector, with and without supersymmetry, and perform
a dedicated signal-to-background analysis
to test the scope of this channel 
for the LHC running at the design energy (14 TeV), for
300\,fb$^{-1}$ (standard) and 3000\,fb$^{-1}$ (high) integrated
luminosities. We find that, while this channel
does not show much promise for a supersymmetric \hpm\ state,  
significant portions of the parameter
spaces of several two-Higgs doublet models are testable. 

\end{abstract}
\vskip 20mm
\noindent {\footnotesize $^\dagger$E-mails:\\
{\tt 
{Rikard.Enberg@physics.uu.se},\\
{William.Klemm@physics.uu.se},\\
{S.Moretti@soton.ac.uk},\\
{S.Munir@apctp.org},\\
{Glenn.Wouda@physics.uu.se}.
}
}

\newpage
\section{\label{intro}Introduction}

A charged Higgs boson, $\hpm$, is predicted in many models of new physics,
with and without Supersymmetry (SUSY). The observation of a
$\hpm$ at the Large Hadron Collider (LHC) is thus expected to provide
concrete evidence of physics beyond the Standard Model (SM). The
strategies for such searches depend on the mass, $\mhpm$, of the
charged Higgs boson. A $\hpm$ lighter than
the top quark can be produced in  $t\rightarrow H^+ b$ and
$\bar{t} \rightarrow H^-\bar{b}$ decays, where the top quarks are
produced in pairs in $q\bar{q}$
annihilation and $gg$ fusion (see\cite{Djouadi:2005gj} and
references therein). When $\mhpm > m_t - m_b$, $bg\rightarrow tH^-$ and $gg\rightarrow tH^-\bar{b}$ are by far the dominant production
processes.\footnote{These are in fact one and the same process,
  describing the underlying dynamics in two different regimes, when
  combined with the parton distribution functions (pdfs). A
  combination of these two modes with a subtraction of the common
  terms is the preferred computational method, as described originally
  in\cite{Maltoni:2003pn,Boos:2003yi} for neutral Higgs boson production and adapted later in\cite{Moretti:2002ht,Assamagan:2004gv}
  for charged Higgs boson production, with an implementation of the
 latter made available
  in\cite{Alwall:2004xw,Alwall:2003tc}. (Also, see
  Refs.\cite{Zhu:2001nt,Plehn:2002vy} for a discussion on the
  QCD accuracy at the next-to-leading order (NLO).) 
Further aspects in this context relevant to our analysis can be found
in \refsec{signal} below.} As for the decays, $H^\pm \rightarrow
\tau\nu$ \footnote{We do not
  distinguish between fermions and anti-fermions when their identity
  is either unspecified or can be inferred from the context.} 
 is the dominant mode as long as $\mhpm <
m_t + m_b $, beyond which $H^\pm \rightarrow tb$ becomes
the leading decay channel with branching ratio (BR) approaching
unity.

The Minimal Supersymmetric Standard Model (MSSM) is an example of a 
scenario  predicting charged Higgs states.
In fact, it contains a
total of five physical Higgs states. Among the neutral ones
are included two CP-even states, with the lighter one denoted by $h$
and the heavier by $H$, a CP-odd state, $A$,
and  there is also a charged pair \hpm. The detection of
an MSSM $\hpm$ lighter than the top quark is rather
straightforward for a wide range of \tanb\ (where $\tanb\equiv
v_2 /v_1$, with $v_1$ and $v_2$ being the vacuum expectation values
(VEVs) of the two Higgs doublet fields $\Phi_1$ and $\Phi_2$).
$H^\pm \rightarrow \tau\nu$ is the dominant decay mode of such a \hpm\
for all $\tan\beta$. For $\mhpm > m_t + m_b$, the large reducible and 
irreducible backgrounds make the search for
$\hpm$ in the $tb$ decay mode notoriously
difficult\cite{Gunion:1986pe} (see \cite{Assamagan:2004tt,Lowette:2006bs}
for experimental simulations). 
However, some studies\cite{Barger:1993th,*Gunion:1993sv,
Miller:1999bm,*Moretti:1999bw} concluded that
the LHC discovery potential of a $\hpm$ state with mass $\lesssim
600$\,GeV is satisfactory in this decay
channel, but only for very small, $\lesssim 1.5$, or very large,
$\gtrsim 30$, values of $\tan\beta$. 
It has also been shown\cite{Odagiri:1999fs,*Raychaudhuri:1995cc} that
the $H^\pm \rightarrow \tau\nu$ decay mode can be used at the LHC even for
$200\gev < \mhpm < 1$\tev\ provided $\tan\beta
\gtrsim 3$. In fact, if the distinctive
$\tau$-polarisation\cite{Bullock:1991fd,*Bullock:1992yt} is used,
the $H^\pm \rightarrow \tau\nu$ channel 
 can provide at least as good a heavy $H^\pm$ signature as the 
$H^\pm\rightarrow tb$ decay mode (for the large $\tan\beta$
regime\cite{Roy:1999xw}). 

At the LHC several searches have been carried
 out for {\hpm}'s lighter as well as heavier than the top quark. 
The CMS collaboration has recently released exclusion limits\cite{CMS:2014pea}
for a \hpm\ lying in the 180\gev\ -- 600\gev\ mass range. 
That study assumes $gg \rightarrow tH^-\bar{b}$ production 
and $H^\pm \rightarrow tb$ and
$H^\pm \rightarrow \tau\nu$ decay modes and is based on 19.7\,fb$^{-1}$ of data
collected at $\sqrt{s}=8$\tev. An earlier analysis\cite{CMS:2014cdp}
based on the same dataset provided exclusion limits in the $H^\pm
\rightarrow \tau\nu$ decay channel for 
$80\gev < \mhpm< 160\gev$, assuming $t\bar{t} \rightarrow H^\pm W^\pm
b\bar{b}$ production, and for $180\gev < \mhpm< 600\gev$, using the 
inclusive $pp \rightarrow tH^-(b)$ production mode. 
The same production and decay modes have also been analysed 
 by the ATLAS collaboration\cite{hpmATLAS2014} based on 19.5\,fb$^{-1}$ of data at $\sqrt{s}=8$\tev,
providing exclusion limits for $80\gev < \mhpm< 160\gev$ and
$180\gev < \mhpm< 1$\tev. In an earlier ATLAS study\cite{Aad:2013hla}
based on 4.7\,fb$^{-1}$ of data at $\sqrt{s}=7$\tev, the $H^\pm \rightarrow
cs$ decay channel has also been probed for \hpm\ lying in the mass range 90\gev\ --
150\gev.

Note, however, that the two dominant decay channels mentioned above, i.e., $tb$ and
$\tau\nu$, leave the $1.5 \lesssim \tan\beta \lesssim 3$ window
virtually unexplorable for a \hpm\ heavier than the top quark in the MSSM. 
Importantly, it is for such small values of
$\tan\beta$ that the BR($H^\pm\rightarrow W^\pm h $) becomes sizeable, reaching
the percent level.
The detectability of a Supersymmetric $H^\pm$ in the $W^\pm h$ decay channel  was
studied in\cite{Drees:1999sb,*Moretti:2000yg}, where
it was noted that a $H^\pm$ with mass around 200\gev\ could be
detectable at the LHC with $\sqrt{s}=14$\,TeV and $\mathcal{L}=300$ fb$^{-1}$,
for $\tanb = 2 - 3$. But there are two caveats.
First, in these studies the mass of $h$ was not
fixed to the value eventually measured at the LHC. Second,
such low values of \tanb\ may at first glance appear to be
excluded by the LEP2 Higgs boson searches\cite{Sopczak:2000pu}, particularly for 
low $m_A \sim 100$\gev. However, as discussed
in\cite{Djouadi:2013vqa}, the LEP limit
typically assumes a SUSY-breaking scale, $M_{\rm SUSY}$, in the vicinity
of 1\tev, which should be relaxed
owing to the fact that SUSY remains undiscovered, implying a
significantly higher breaking scale. Now, a realistic SUSY model ought to
contain a Higgs boson, \hobs, consistent with the one discovered at
the LHC\cite{Chatrchyan:2012ufa,*Aad:2012tfa} and hence satisfying 
the `observational constraint,' $122\gev\lesssim m_{\hobs} \lesssim
128\gev$, which supersedes the LEP limit. The large allowed mass
window is to take into account
the theoretical uncertainties in the calculation of the \hobs\
mass in the model. All such aspects clearly need to be re-assessed
in light of the latest experimental results.

Besides the above observational constraint on the mass of the Higgs boson, the
LHC measurements of its signal strengths in various production and decay
channels also strongly constrain the  parameter space of the MSSM wherein a 
$H^\pm$, potentially visible via the $W^\pm  \hobs$ decay, can be
obtained. In its singlet-extension, the Next-to-Minimal Supersymmetric 
Standard Model (NMSSM), 
the mass of the SM-like Higgs boson satisfying the mentioned mass
constraint can be achieved in a more natural way, without requiring
large radiative corrections from the
stop sector. Such a Higgs boson, in fact, favours a lighter $H^\pm$, as
we shall discuss in detail below. Moreover, in this model, which
contains a total of 5 neutral Higgs
states, the role of $\hobs$ can be played by
the any of the two lightest CP-even Higgs bosons, $H_1$ or $H_2$, 
alternatively\cite{Ellwanger:2011aa,*King:2012is,*Cao:2012fz,*Ellwanger:2012ke,*Gherghetta:2012gb}.

If one leaves aside SUSY, one of the simplest non-trivial
extensions of the SM is represented by a 
2-Higgs doublet model (2HDM), which contains two Higgs 
doublets with different Yukawa assignments (see\cite{Branco:2011iw}
for a review). Notably, this structure (albeit limited to one specific Yukawa configuration) is necessary in the MSSM, implying that 
the Higgs spectrum in a CP-conserving 2HDM is the same as in the
MSSM, containing three neutral Higgs bosons and a charged
  pair. However,
the absence of SUSY relations amongst the Higgs boson masses 
allows much more
  freedom to alternatively identify the discovered SM-like Higgs
state with either of the two CP-even Higgs bosons of a
2HDM. Depending on the way the Higgs doublets are assigned
charges under a $Z_2$ symmetry imposed in order to avoid large
 flavour-changing neutral currents (FCNCs), the 2HDMs are generally
 divided into four different types. In the `aligned'
 2HDM\cite{Pich:2009sp} (A2HDM), instead of the $Z_2$ symmetry, 
a Yukawa-alignment is enforced in order to prevent large FCNCs.

From the point of view of \hpm\ searches, results
obtained in the MSSM can be easily translated to the case of a 2HDM Type II,
as long as SUSY states are very heavy, i.e.,
decoupled\cite{Bisset:2000ud,*Bisset:2003ix}. 
This is somewhat more involved in the case of the other three
ordinary Types and the A2HDM, although still possible
 (see\cite{Belyaev:2009nh,*Aoki:2011wd} and\cite{Akeroyd:2012yg,*Akeroyd:2014cma}, 
respectively). Some dedicated analyses of the 2HDMs to constrain them 
using the latest data from the LHC have also been
performed recently\cite{Dev:2014yca,*Broggio:2014mna}. The key
phenomenological difference in the 2HDMs from the SUSY models in
general, and the MSSM and NMSSM in particular, is that
there are no light SUSY particles to provide cancellations (induced by the different spin statistics between SM and SUSY states) in low energy
observables, chiefly from flavour dynamics. It is in fact the latter
(e.g., limits on the $Z\to b\overline b$ and $b\to s\gamma$ decays) that generally
produce severe constraints on the mass of \hpm\ in the standard 2HDMs,
pushing it to be larger than the top quark 
mass\cite{Hou:1987kf,*Rizzo:1987km,*Grinstein:1987vj,*Grinstein:1990tj,*Borzumati:1998tg,*Borzumati:1998nx,*Misiak:2006zs}.
In the A2HDM, however, one can obtain $m_{H^\pm}< m_t$ in a viable
region of the parameter space\cite{Jung:2010ik,*Jung:2010ab,*Jung:2012vu,*Cree:2011uy}.  

In this article we analyse the possibility of establishing a
 $H^\pm\to W^\pm \hobs$ signal in the next LHC run in all the models 
mentioned above, which are those where some relevance of such
a decay has been established in the literature previously. We exploit the requirement  on 
$\hobs$ to have a mass around 125\gev, so that the \mhpm\ range accessible via this signature starts at about 200 GeV and 
extends to nearly 500 GeV, as for heavier masses the $tH^\pm$ production cross section becomes too low. We first discuss the consistency
of the corresponding regions of the parameter spaces of these models with the current 
Higgs boson data from the LHC.  We further assess the effects of
imposing constraints from $b$-physics and, in the case of SUSY models,
cold dark matter (DM) relic density measurements. We also carry out
a model-independent detector-level analysis of the expected
LHC sensitivity in the $\hpm \rightarrow W^\pm \hobs $ channel
with $\sqrt{s} = 14$\tev. In doing so, we exploit the
 knowledge of the mass of \hobs, which will result in a substantial improvement
in the efficiency of previously advocated\cite{Drees:1999sb,*Moretti:2000yg} 
kinematical selections for
the extraction of the signature of concern here, which we use for
guidance. We then compare the sensitivities expected for various
integrated luminosities at the LHC with the cross sections obtainable 
for this channel in each model
considered in the presence of the aforementioned experimental constraints.\footnote{See\cite{Coleppa:2014cca} for a similar analysis
  for some Type II 2HDM benchmark points.} It will be the interplay between the improved selection and the reduced parameter space available following
the Higgs boson discovery (with respect to the setups assumed in earlier analyses of the\hpm\ decay mode considered here) that will determine the
actual situation at present.

The article is organised as follows. In \refsec{production} we will
discuss the production and decay mechanisms of the $\hpm$ considered
in our analysis. In \refsec{models}, we
will discuss some salient features of the models analysed. 
In \refsec{method} we will provide some details of the scans of the
parameter spaces of these models and of the experimental
constraints imposed in our study. In
\refsec{signal} we will explain our signal-to-background
analysis. In \refsec{results} we will present our results and in
\refsec{concl} our conclusions.

\section{\label{production} Production and decay of \hpm}

The dominant production process at the LHC for a $\hpm$
heavier than the top
quark is its associated production with a single top, with the
relevant subprocesses being $bg \to tH^-$ and $gg \to t\bar{b} H^-$
 (plus charge conjugated channels).
The division between these two subprocesses is not clear-cut. The $gg$
amplitude can be seen as a tree-level contribution to the
NLO amplitude that includes a virtual
$b$-quark, with the $bg$ process making the LO
amplitude. In the $gg$ process we may view the $b$-quarks (the virtual
$b$ and the emitted $b$) as resulting from a splitting of the gluon
and the corresponding amplitude contains the exact kinematics of this
splitting. In the $bg$ process the $b$-quark instead comes from the
parton distribution of the proton. The $b$-quark is then a collinear
parton arising from a splitting in the evolution of the pdfs. This contribution to the
amplitude contains a collinear approximation of the kinematics and also a resummation of large logarithms in the
factorisation scale that is not present in the $gg$ amplitude.

When calculating the cross section for $pp \to t\hpm + X$ the $bg$ and
$gg$ contributions to the amplitude cannot be added naively because that
would result in double counting between the two contributions. There
is a correct procedure to compute the total cross
section\cite{Borzumati:1999th}, but it does not generalise to
the differential cross section needed for Monte Carlo (MC)
simulations. In Ref.\cite{Alwall:2004xw} a method for event generation
without double counting was introduced, and an add-on, called MATCHIG,
to the event generator Pythia\,6\cite{Sjostrand:2006za} was
constructed. In this framework events are generated both for the $bg$ and
$gg$ processes and for the double-counting contribution. Events
corresponding to the 
double counting have negative weights and should be subtracted from 
the positive weighted $bg$ and $gg$ processes. We have used MATCHIG
in our simulations.\footnote{The process $bg \to tH^-$ already exists in
  the publicly available Pythia package.}

The process $pp \to t\hpm + X$ has also been calculated at NLO and has
been implemented\cite{Klasen:2012wq} in the POWHEG BOX MC
framework\cite{Nason:2004rx, *Frixione:2007vw, *Alioli:2010xd}, which
includes matching to parton showers. At NLO the $bg$ and $gg$
contributions are both part of the amplitude. It has also been
implemented\cite{Weydert:2009vr} in the MC@NLO
framework\cite{Frixione:2002ik}. In\cite{Klasen:2012wq} it was shown
that the MATCHIG program produces very similar kinematical
distributions to the POWHEG implementation except at very large
transverse momentum, $p_T>200$\gev\ of the $t\hpm$ pair. The overall
normalisation is, however, larger for the NLO calculations. The ratio
between the total cross sections at NLO and LO depends on
the model parameters via the mass spectrum, but for an example choice 
of 2HDMs it was found to be around a factor
2 for the Tevatron energies and a factor 1.4 for the LHC
energies\cite{Klasen:2012wq}. We do not consider this NLO enhancement of the signal
in this paper for consistency, as we are only able to simulate the backgrounds at LO, but one should bear in mind that our quoted sensitivities may be
somewhat stronger if NLO effects were systematically taken into account. 

The spin/colour summed/averaged squared amplitude for the $gb \rightarrow tH^-$ production process is given by\cite{Moretti:2002eu}
\begin{eqnarray}
\label{eq:sigma}
\overline{|\mathcal{M}|^2}
=\frac{g^2_{q\hpm}}{2m_W^2} \, \frac{g_s^2 g_2^2}{4N_c}|V_{tb}|^2 \,  
\frac{(u-\mhpm^2)^2}{s(m_t^2-t)}\left[1+2\frac{\mhpm^2-m_t^2}{u-\mhpm^2}\left(1+\frac{m_t^2}{t-m_t^2}+\frac{\mhpm^2}{u-\mhpm^2}\right)\right],
\end{eqnarray}
where $g_s$ and $g_2$ are the $SU(3)_C$ and $SU(2)_L$ gauge couplings, $N_C=3$ is
the number of colours and $V_{tb}$ is the relevant CKM matrix element.
See Refs.\cite{Miller:1999bm} and\cite{Guchait:2001pi} for the $gg\to
tH^-\bar b$ amplitudes and graphs. The total cross section is
proportional to the coupling $g^2_{q\hpm}$, as noted in the equation above, which is 
the only model dependent factor for a given \mhpm.
This factor depends on the masses, $m_t$ and $m_b$, of the $t$ and $b$
quarks, respectively, as well as the parameter \tanb, and will be
discussed in the next section for each model considered here. 
As shown in\cite{Alwall:2004xw}, the total
cross section for a charged Higgs mass above $m_t$ is actually
well-approximated by the $bg$ cross section. However, since the $bg$
and the $gg$ contributions lead to different kinematical distributions in
the MC simulations, as noted above, we included both these
contributions in our MC simulations. 

Finally, as noted in the Introduction, this study aims to
  exploit the $H^\pm \rightarrow W^\pm \hobs$ decay channel at the LHC. Of
relevance for this particular process is the coupling of \hpm\ to a
generic neutral Higgs boson, $H_i$, and the $W$ boson, given by 
\begin{equation}
\label{eq:hpmhwcpl}
g_{H_i H^+ W^-} = \frac{g_2}{2}(\cos\beta S_{i2} - \sin\beta S_{i1})  \,,
\end{equation}
where $S_{i1}$ and $S_{i2}$ are the elements of the mixing matrix 
that diagonalises the CP-even Higgs mass matrix in the
model. It is clear that this coupling depends strongly on
  \tanb, both explicitly and through the elements $S_{i1}$ and
  $S_{i2}$, (except in the A2HDM, as will be explained later) making
the $H^\pm \rightarrow W^\pm \hobs$ decay process highly sensitive to
this parameter.

\section{\label{models} The models}

\subsection{Supersymmetric models}

The Supersymmetric models considered here contain two Higgs doublets, $\Phi_1$ and
$\Phi_2$, which make the scalar components of the
superfields $\widehat{H}_d$ and $\widehat{H}_u$, respectively. The  
field $\Phi_1$ is needed for generating the masses
of the $d$-type quarks and leptons and $\Phi_2$ those of the $u$-type quarks.
The coupling of the charged Higgs boson to the quarks, defined in
Eq.\,(\ref{eq:sigma}) as the factor $g^2_{q\hpm}$, is given in these models as
\begin{equation}
\label{eq:gqhpm}
g^2_{q\hpm} = m_b^2\tan^2\beta + m_t^2\cot^2\beta\,.
\end{equation}
Thus the amplitude for the $gb \rightarrow tH^-$ process is
maximal for either small or large \tanb.  \\

\noindent $\bullet$ \textbf{MSSM}\\[0.15cm]
\noindent
The MSSM Superpotential, from which the scalar potential is derived, is
given as
\begin{equation}
\label{eq:MSSMsuper}
W_{\rm MSSM} = h_u\, \widehat{Q} \cdot \widehat{H}_u\;
\widehat{U}^c_R \: + \: h_d\, \widehat{H}_d \cdot \widehat{Q}\;
\widehat{D}^c_R \: + \: h_e\, \widehat{H}_d \cdot \widehat{L}\;
\widehat{E}_R^c \: +\: \mu \widehat{H}_u \cdot \widehat{H}_d \,,
\end{equation}
where $\widehat{Q},\,\widehat{U}_R^c,\,\widehat{D}_R,\,\widehat{L}$ and
$\widehat{E}_R$ are the quark and lepton superfields and $h_u,\,h_d$
and $h_e$ are the corresponding Yukawa couplings. In this model, the mass of $H^\pm$ is given at LO as
\begin{equation}
\label{eq:MSSMmhpm}
\mhpm^2 = m_A^2 + m_W^2\,,
\end{equation}
where $m_W$ is the mass of the $W$ boson. In order to allow the $\hpm
\rightarrow W^\pm \hobs$ decay, one requires $\mhpm > m_{H_{\rm obs}}
+ m_W$, which translates into the requirement $m_A
\gtrsim 190$\gev. In the MSSM, under such a condition, the tree-level
mass of the SM-like Higgs boson, \hsm, has an upper limit
\begin{equation}
\label{eq:MSSMlimit}
\mhsm^2 \leq m^2_Z \cos^2 2\beta\,,
\end{equation}
where $m_Z$ is the mass of the $Z$ boson. Therefore, if the \hsm\ is identified
with the \hobs\ and hence required to have a mass close to 125\gev\ 
in accordance with the LHC measurement, a large value of \tanb\ is necessary.
Furthermore, the absence of any significant deviations
of the signal strengths of the \hobs\ from the SM
expectations so far\cite{CMS-H-twiki,*ATLAS-H-twiki}
seems to be pushing the MSSM towards the so-called `decoupling
regime'. This regime corresponds to $m_A \gtrsim
150$\gev\ for $\tanb \gtrsim 10$ and yields SM-like couplings of the
\hsm, in addition to a maximal tree-level mass, as noted above. 
The net effect of all these observations is that 
a \hpm\ with mass greater than 200\,GeV and a \hsm\ with the correct
mass and SM-like couplings can be obtained simultaneously only for large 
 \tanb. However, according to Eqs.\,(\ref{eq:hpmhwcpl}) and
 (\ref{eq:gqhpm}), $\tanb\sim 10$ not only diminishes
 the BR($H^\pm \rightarrow W^\pm \hsm$) but also the 
$gb\rightarrow tH^-$ cross section. 

The complete MSSM contains more than 120 free parameters in
addition to those of the SM. In its phenomenological version, the pMSSM, one 
assumes the matrices for the sfermion
masses and for the trilinear scalar couplings to be diagonal, which 
reduces the parameter space of the model considerably. 
Here, since we are
mainly concerned with the Higgs sector of the model, we further impose
the following mSUGRA-inspired (where mSUGRA stands for minimal
supergravity) universality conditions: 
$$m_0 \equiv M_{Q_{1,2,3}} = M_{U_{1,2,3}} = M_{D_{1,2,3}} =
M_{L_{1,2,3}} = M_{E_{1,2,3}}\,,$$
$$m_{1/2} \equiv 2M_1 = M_2 =  \frac{1}{3} M_3\,,$$
\begin{equation}
A_0 \equiv A_t = A_b = A_{\tau}\,,
\end{equation}
 where $M_{Q_{1,2,3}},\,M_{U_{1,2,3}},\,M_{D_{1,2,3}}
 ,\,M_{L_{1,2,3}}$ and $M_{E_{1,2,3}}$ are the soft masses of the
 sfermions, $M_{1,2,3}$ those of the gauginos and $A_{t,b,\tau}$ the
 soft trilinear couplings. This leaves us with a total of
six free parameters, namely $m_0$, $m_{1/2}$, $A_0$, $m_A$, \tanb\ and
the Higgs-higgsino mass parameter $\mu$. \\

\noindent $\bullet$ \textbf{NMSSM}\\[0.15cm]
\noindent
The NMSSM\cite{Fayet:1974pd,*Ellis:1988er,Durand:1988rg,*Drees:1988fc,Miller:2003ay}
(see, e.g., \cite{Ellwanger:2009dp,Maniatis:2009re} for reviews) 
 contains a singlet Higgs field in addition to the two doublet fields of the MSSM. 
The scale-invariant Superpotential of the NMSSM is written as
\begin{equation}
\label{eq:superpot}
W_{\rm NMSSM}\ =\ {\rm MSSM\;Yukawa\;terms} \: + \: 
\lambda \widehat{S} \widehat{H}_u \cdot \widehat{H}_d \: + \:
\frac{\kappa}{3}\ \widehat{S}^3\,,
\end{equation}
where $\widehat{S}$ is the additional Higgs singlet Superfield and
$\lambda$ and $\kappa$ are dimensionless Yukawa couplings. The
introduction of the new singlet field results in a total of five
neutral Higgs mass eigenstates and a $H^\pm$ pair, after rotating away 
the Goldstone bosons. In the NMSSM, the
MSSM upper limit on the tree-level mass of the SM-like Higgs boson,
given in Eq.\,(\ref{eq:MSSMlimit}), gets modified as
\begin{equation}
\label{eq:mhlim}
\mhsm^2 \leq m_Z^2\cos^2 2\beta 
 + \frac{\lambda^2 v^2 \sin^2 2\beta}{2} -
  \frac{\lambda^2 v^2}{2\kappa^2}\left[\lambda -\sin2\beta \left(\kappa +
  \frac{A_\lambda} {\sqrt{2}s}\right)\right]^2 \,,
\end{equation}
where $v\equiv\sqrt{v_1^2 + v_2^2}=246$\gev, 
$s$ is the VEV of the singlet field
and $A_\lambda$ is the soft SUSY-breaking parameter corresponding to
the coupling \lam. 
Clearly, for large values of \lam\ and small \tanb, the second term 
in the above equation gives a significant positive contribution to the \hsm\ mass.

The mass expression for $H^\pm$ in the NMSSM is given as
\begin{equation}
\label{eq:NMSSMmhpm}
\mhpm^2 = m_A^2 + m_W^2 - \frac{v^2\lambda^2}{2}\,,
\end{equation}
where $m_A^2$ is, in contrast with the MSSM, the diagonal entry $[M^2_A]_{11}$ of the pseudoscalar mass matrix $M^2_A$ of the model, given by 
\begin{equation}
 m_A^2 = [M^2_A]_{11} = \frac{\sqrt{2}\lambda s}{\sin2\beta}(A_\lambda + \frac{\kappa
   s}{ \sqrt{2}})\,.
\label{eq:ma2}
\end{equation}
 Again, for a given value of \tanb, the negative third term in
 Eq.\,(\ref{eq:NMSSMmhpm}) results in a smaller
$\mhpm^2$ in the NMSSM compared to that in the
MSSM, where it is given by the first two terms only. This negative
contribution increases with the size of \lam. 

A crucial observation here is that a large 
 \lam, necessary to obtain sufficiently small \mhpm,
 has the dual advantage of enhancing also the tree-level mass of \hsm, as
 noted above. Such a scenario
is therefore more natural than the one with a very MSSM-like \hsm, since
a much smaller amount of fine-tuning is 
required to achieve the correct Higgs boson mass via radiative
corrections. But large $\lambda$ also implies a substantial singlet component in 
\hsm, which could result in significantly reducing its
couplings to fermions and gauge bosons compared to those of the SM Higgs
boson. However, recent
studies\cite{Ellwanger:2011aa} have shown that, for large
$\lambda$ and small $\tan\beta$, the $\hsm$ of the model, which can
correspond to either $H_1$ or $H_2$, can
still be consistent with the LHC Higgs boson data. The signal
strength of $\hsm$ in the $\gamma\gamma$ decay channel in such a
scenario can in fact be much larger than that of a SM-like Higgs boson, 
owing to a reduction in the BR($\hsm\rightarrow b\bar{b}$) compared to the true SM case. 
We point out here that, as in the MSSM, the \hsm\ in the
  NMSSM will also be identified with \hobs, since it is
  assumed to be the Higgs boson observed at the LHC.

The phenomenological version of the NMSSM that we study here contains three new
parameters in addition to those of the pMSSM, mentioned earlier, with
$\mu$ replaced by \mueff ($\equiv \lam s$) and $m_A$ traded for \alam. 
These include \lam, \kap\ and \akap, the latter being a
dimensionful coupling originating in the SUSY-breaking 
part of the Higgs potential. 

\subsection{2HDMs}

A generic non-Supersymmetric 2HDM is defined by 
its scalar potential and its Yukawa couplings. The two Higgs doublets
in such a model are written in terms of their VEVs and the physical Higgs states as
\begin{equation}
\Phi_1=\frac{1}{\sqrt{2}}\left(\begin{array}{c}
\displaystyle \sqrt{2}\left(G^+\cos\beta -H^+\sin\beta\right)  \\
\displaystyle v_1-h\sin\alpha+H\cos\alpha+\mathrm{i}\left( G\cos\beta-A\sin\beta \right)
\end{array}
\right),
\end{equation}
\begin{equation}
\Phi_2=\frac{1}{\sqrt{2}}\left(\begin{array}{c}
\displaystyle \sqrt{2}\left(G^+\sin\beta +H^+\cos\beta\right)  \\
\displaystyle v_2+h\cos\alpha+H\sin\alpha+\mathrm{i}\left( G\sin\beta+A\cos\beta \right)
\end{array}
\right),
\end{equation}
where $\alpha$ is the mixing angle of the two CP-even Higgs bosons,
\tanb\ has been defined earlier and $G$ and $G^+$ are the Goldstone
bosons. The most general, CP-conserving potential for two Higgs doublets reads
\begin{equation}
\begin{split}
\mathcal{V}_{\text{2HDM}}  &= m_{11}^2\Phi_1^\dagger\Phi_1+ m_{22}^2\Phi_2^\dagger\Phi_2
-[m_{12}^2\Phi_1^\dagger\Phi_2+ \, \text{h.c.} ] \\
& +\half\lambda_1(\Phi_1^\dagger\Phi_1)^2
+\half\lambda_2(\Phi_2^\dagger\Phi_2)^2
+\lambda_3(\Phi_1^\dagger\Phi_1)(\Phi_2^\dagger\Phi_2)
+\lambda_4(\Phi_1^\dagger\Phi_2)(\Phi_2^\dagger\Phi_1) \\
& +\left\{\half\lambda_5(\Phi_1^\dagger\Phi_2)^2
+\big[\lambda_6(\Phi_1^\dagger\Phi_1)
+\lambda_7(\Phi_2^\dagger\Phi_2)\big]
\Phi_1^\dagger\Phi_2+\, \text{h.c.}\right\}\,.
\label{eq:2hdmpot}
\end{split}
\end{equation}
Through the minimisation conditions of the Higgs potential
  above, $m_{11}^2$ and $m_{22}^2$ can be traded for the VEVs $v_1$
  and $v_2$, respectively. Furthermore, the tree-level mass relations allow the
  quartic coupling $\lambda_{1-5}$ in Eq.\,(\ref{eq:2hdmpot}) to be
  substituted by the four physical Higgs boson masses and the neutral
  sector mixing parameter $\sin(\beta-\alpha)$.
Thus, in contrast with the SUSY models, in the 2HDMs the masses of the 
Higgs bosons are free input parameters, along with
$\lambda_6,\,\lambda_7 ,\,m_{12}^2,\,\sin(\beta-\alpha)$ and \tanb.

In the 2HDMs, the Yukawa couplings of the fermions
are also {\it a priori} free parameters. However, depending on how the two 
Higgs doublets couple to the fermions, FCNCs can be mediated by scalars at
the tree level. The requirement of no large FCNCs thus puts very strong
restrictions on the coupling matrices. There are two general
approaches for avoiding large FCNCs. One way is to impose a
$Z_2$ symmetry so that each type of fermion only couples to one of the
doublets (``natural flavour conservation'')\cite{Glashow:1976nt,
  Paschos:1976ay}. The same symmetry then holds also in the scalar
potential (forcing $\lambda_6 = \lambda_7 = 0$), up to the soft
breaking terms with parameter $m_{12}^ 2$, thus further reducing the
number of free parameters.

As noted in the Introduction, there are four ways of assigning the $Z_2$ charges, giving
2HDMs of Types I, II, X and Y. One defines as Type I the model where
only the doublet $\Phi_2$ couples to all fermions; Type II is the
scenario similar to the MSSM, where $\Phi_2$ couples to up-type quarks 
and $\Phi_1$ couples to down-type quarks
and leptons; in a Type X (or Type IV or `lepton-specific') model $\Phi_2$ couples
to all quarks and $\Phi_1$ couples to all leptons; and a Type Y (or Type III or
`flipped') model is built such that $\Phi_2$ couples to up-type quarks
and to leptons and $\Phi_1$ couples to down-type quarks. 
The Type X and Type Y models 
have a similar phenomenology to Type I and II, respectively,
especially in the context of this study. Specifically, $g^2_{q\hpm}$
is the same in the Type I
  and Type X models. Similarly, the Type Y model has a similar Yukawa
  structure, and consequently $g^2_{q\hpm}$, as Type II, except for
  the leptons which couple to a different
  Higgs doublet in either of the two models. This, incidentally, implies that there
  is no \tanb-enhancement in the Type Y model to affect the
  BR($\hpm\to\tau\nu$). We therefore
consider only the Type I and Type II models, referred to as 2HDM-I and
2HDM-II, respectively, which are the most well-known ones.

Another way to achieve small FCNCs without imposing natural flavour
conservation is to postulate that the Yukawa
coupling matrices of the two Higgs doublets are proportional to each
other, i.e., they are aligned. This approach has been adopted in the aforementioned
A2HDM\cite{Pich:2009sp}, where both scalar doublets 
($\Phi_1$ and $\Phi_2$) couple to all types of fermions. In the
  $Z_2$-symmetric 2HDMs discussed above the Yukawa couplings
  are determined solely by the parameter \tanb, while the
  CP-conserving A2HDM instead has separate parameters for the up-type
  quarks, the down-type quarks and the leptons, usually denoted by 
$\beta^U$, $\beta^D$ and $\beta^L$. In the A2HDM there is no specific
basis singled out by the fermionic sector due to the absence
of the $Z_2$ symmetry. For this study we choose the basis where
only one doublet acquires a VEV, called the `Higgs basis'. In this basis
the input parameters include $\sin \alpha$ (where $\alpha$ is the
angle that diagonalises the CP-even Higgs-sector), $\lambda_2,\,
\lambda_3,\,\lambda_7$ and the above-mentioned alignment angles 
$\beta^{U,D,L}$, in addition to the physical Higgs boson masses. 

\begin{table}[ht!]
\begin{center}
\begin{tabular}{|c|c|c|c|}
\hline
 & 2HDM-I & 2HDM-II & A2HDM \\
\hline
 $g^2_{q\hpm}$ & $m_b^2\cot^2\beta + m_t^2\cot^2\beta $ &   $ m_b^2\tan^2\beta + m_t^2\cot^2\beta $ & $ m_b^2\tan^2\beta^D + m_t^2\tan^2\beta^U $ \\
\hline
\end{tabular}
\end{center}
\caption{The expressions for $g^2_{q\hpm}$ in the different 2HDMs
  considered in this paper.}
\label{table:g2qHm}
\end{table}

The expressions for $g^2_{q\hpm}$ in Eq.~(\ref{eq:sigma}) for the
different 2HDMs  (including the A2HDM) are given in Table~\ref{table:g2qHm}. It should
be noted that $g^2_{q\hpm}$ in the 2HDM-II is identical to the
one in the SUSY models.

\section{\label{method} Model scans and experimental constraints}

We have performed scans of the parameter spaces of all the models
considered here, requiring \mhpm\ to lie in the 
200\gev\ --500\gev\ range. For each scenario except the MSSM, 
we carried out two separate scans for the cases with $H_1$ and $H_2$
alternatively playing the role of \hobs,
i.e., having mass
near 125\gev\ and SM-like signal rates in the $\gamma\gamma$
and $ZZ$ decay channels. We point out here that in the
MSSM it is not possible to obtain a $H$ with a mass around 125\gev\
while also requiring $\mhpm \gtrsim 200$\gev, as their masses lie
very close to each other by theoretical construction. In the case of
the SUSY models, since the masses of the
scalar Higgs bosons are derived and not input parameters, we used the nested 
sampling package MultiNest-v2.18\cite{Feroz:2008xx} for efficiently scanning their
parameter spaces. 

The mass spectra and Higgs boson decay BRs for each scanned
point of the MSSM, the NMSSM and the 2HDMs were computed using the public
packages SUSY-HIT-v1.3\cite{Djouadi:2006bz}, \nmssmtools
-v4.2.1\cite{NMSSMTools} and 2HDMC\cite{Eriksson:2009ws}, respectively. 
For a point to be accepted in a given scan, it had to 
pass the condition $122\gev \leq m_{\hobs} \leq 128\gev$ for
the SUSY models and $123\gev \leq m_{\hobs} \leq 127\gev$ in the
2HDMs. This is to take into account the experimental as well theoretical
uncertainties (which are understandably larger in the presence of
SUSY) in $m_{\hobs}$ predicted in the two scenarios.
As for the $b$-physics observables, the points
  for which their theoretically evaluated values did not lie in the
  following ranges were rejected during the scans for the NMSSM
    and the A2HDM.
 \begin{itemize}
\item $2.63 \times 10^{-4} \leq \brbxsgamma \leq 4.23 \times 10^{-4}$,
\item $0.71 \times 10^{-4} < \brbutaunu < 2.57 \times 10^{-4}$,
\item  $1.3 \times 10^{-9} < \brbsmumu < 4.5 \times 10^{-9}$.
\end{itemize}
These 95\% confidence level ranges are the ones suggested
in the manual of the package \superiso-v3.4\cite{superiso}, which was 
used for the theoretical evaluation of these observables. Additionally, the scan
points were also required to satisfy the constraint $\Delta M_{B_d} = 
(0.507\pm 0.004)$\,ps$^{-1}$, which is based
on\cite{Mahmoudi:2009zx}. In
  the case of the $Z_2$-symmetric 2HDMs, their parameter spaces 
consistent with the $b$-physics
  constraints were adopted directly from\cite{Mahmoudi:2009zx}, so
that these constraints were not tested against during the scans.
Moreover, for SUSY models the (lightest) neutralino DM relic 
density was calculated for every
point using the package \micromegas -v2.4.5\cite{micromegas}. 
Only points with $\Omega_\chi h^2 < 0.131$, assuming a +10\%
theoretical error on the central value of 0.119 measured by
the PLANCK collaboration\cite{Ade:2013zuv}, were retained.

Finally, we used the public package HiggsBounds-v4.1.3\cite{Bechtle:2008jh,
*Bechtle:2011sb,*Bechtle:2013gu,*Bechtle:2013wla}
to test the neutral Higgs bosons other than the \hobs\ in a given case for each
model against the exclusion limits from the Large Electron--Positron
(LEP) collider, the Tevatron and the LHC. This
program also takes care of the exclusion constraints on \hpm\ from the various
LHC searches mentioned in the Introduction.
Finally, the magnitude of a possible Higgs boson signal at the LHC is 
characterised by the signal strength modifier, defined as 
\begin{equation}
\mu^X =  \frac{\sigma(pp\rightarrow H_{\rm obs} \rightarrow
  X)}{\sigma(pp\rightarrow h_{\rm SM} \rightarrow X)}\,,
\end{equation}
where $X$ denotes the decay channel under consideration and
  $h_{\rm SM}$ denotes a 125\gev\ SM Higgs boson. 
The theoretical counterparts of $\mu^X$, which we refer
to as $R^X$ here, 
were obtained from the program HiggsSignals-v1.20\cite{Bechtle:2013xfa} for  
 $X= \gamma\gamma,\,ZZ$.\footnote{The $\gamma\gamma$ and $ZZ$ 
decay channels remain the only ones so far where a 5$\sigma$
excess has been established at the LHC.} In our analysis 
below, while we will show all the good points from our scans, 
we will  highlight the points for which $R^{\gamma\gamma,
  ZZ}$ are consistent with the measured $\mu^{\gamma\gamma, ZZ}$ at
the LHC. The latest publicly available measurements read
\begin{equation}
\mu^{\gamma \gamma} = 1.13 \pm 0.24 ~{\rm and}~\mu^{ZZ} = 1.0 \pm 0.29
\end{equation}
at CMS\cite{CMS-PAS-HIG-14-009} and 
\begin{equation} 
\mu^{\gamma \gamma} = 1.57^{+0.33}_{-0.28} ~{\rm and}~
\mu^{ZZ} = 1.44^{+0.40}_{-0.35}
\label{eq:ATLAS}
\end{equation}
at ATLAS\cite{ATLAS-CONF-2014-009}.\footnote{We note here that the 
ATLAS collaboration has recently made public\cite{Aad:2014eha} an updated measurement, 
$\mu_{\gamma \gamma} = 1.17\pm 0.27$, which is
now comparatively much closer to the SM prediction. However, no updates on
$\mu^{ZZ}$ for the same data set have been released. This implies that even if
we use the newly released $\mu^{\gamma\gamma}$ value, the older and
larger value of $\mu^{ZZ}$ in Eq.\,(\ref{eq:ATLAS}) will still rule out the
corresponding model points, since $R^{ZZ}$ is generally smaller than 
$R^{\gamma\gamma}$.}

\section{\label{signal} Signal and background analysis}

In addition to constraining the parameter spaces of the new
 physics models, knowledge of the mass of $\hobs$ also provides an
additional handle in identifying the $H^\pm\to W^\pm \hobs$ decay. We
focus here on the decay $\hobs\to b\bar{b}$, as it generally has a  
substantial BR and allows for a full reconstruction of
$\hobs$.\footnote{This channel was also recently studied
    in\cite{Coleppa:2014cca}, where it was noted that especially when
    uncertainties become dominated by systematics, the decay
    $\hobs\to\tau^+\tau^-$ can become more relevant due to its smaller 
backgrounds, despite a smaller BR and additional unobservable
neutrinos. In this study, we consider only statistical uncertainties.}
    In particular, we look for the
production channel $pp\to t(b)H^\pm \to W^\mp b (b) W^\pm \hobs$, which,
after semi-leptonic decays of the two $W$ bosons and $\hobs\to
b\bar{b}$, gives a final state of
$bbb(b)jj\ell\nu_\ell$. The main
background for this process is $t\bar{t}$ production, and here we
consider all processes $pp\to t(b)W^\pm b\bar{b}$, where the extra
pair of $b$-quarks can come from the emission of a gluon, a Higgs boson,
or a $Z$. In this section we describe our method for reconstructing the
\hpm\ signal and separating it from the background events to give an
estimate of the sensitivities that could be achieved at the 14\tev\ LHC.

We generate the hard process for the signal using the MATCHIG
package\cite{Alwall:2004xw} with Pythia 6.4.28\cite{Sjostrand:2006za},
thus including the $bg$ and $gg$ contributions and subtracting the
correct double-counting term to get proper $b$-jet momentum
distributions. MadGraph5\_aMC$@$NLO\cite{Alwall:2014hca} is used to
generate the backgrounds. Parton showers and hadronisation for
both signal and background are performed with Pythia
8\cite{Sjostrand:2007gs}, followed by detector simulation with DELPHES
3\cite{deFavereau:2013fsa} using experimental parameters calibrated to
the ATLAS experiment with modified $b$-tagging efficiencies.\footnote{The $b$-tagging used is
  given by $\epsilon_\eta\tanh(0.03 p_T - 0.4)$, with the transverse
  momentum, $p_T$, in GeV, $\epsilon_\eta = 0.7$ for central
  ($|\eta|\leq 1.2$), and $\epsilon_\eta = 0.6$ for forward
  ($1.2\leq|\eta|\leq 2.5$) jets. This choice is a conservative
    one in comparison with the ATLAS high-luminosity projections\cite{ATL-PHYS-PUB-2013-009}.}

For reconstruction and background reduction, we roughly follow the
procedures of previous 
analyses\cite{Drees:1999sb,*Moretti:2000yg}, with the addition of a top veto (described below) to further suppress the background.
\begin{enumerate}
\item Accept events with at least 3 $b$-jets, at least 2 light jets, one lepton ($e$ or $\mu$), and missing energy.  All objects must have transverse momentum $p_T>20$\gev\ and rapidity $|\eta|\leq 2.5$, and must be separated from other objects by $\Delta R > 0.4$.
\item Find a hadronic $W$ candidate from the light jets, taking the pair with the invariant mass $m_{jj}$ closest to $m_W$.  Reject the event if no pair satisfies $|m_{jj}-m_W|\leq 30$\gev.
\item Reconstruct a leptonically decaying $W$ using the lepton and the
  missing energy, by assuming that the missing energy comes entirely from the single neutrino and imposing the invariant mass constraint $m_{\ell\nu}=m_W$.  Because this is a quadratic constraint, there is a two-fold ambiguity in the solution for the longitudinal momentum of the neutrino. If the solutions are real, both are kept, and if they are complex, the real part is kept as a single solution.
\item Apply top veto for high mass searches (``veto first'').
\item Find a Higgs boson candidate from the $b$-jets, taking the pair
  with the invariant mass $m_{bb}$ closest to $m_{H_{\rm obs}} \approx 125$\gev. Reject the event if no pair satisfies $|m_{bb}-m_{H_{\rm obs}}  |\leq 15$\gev.
\item Apply top veto for low mass searches (``veto second'').
\item Reconstruct a top quark using the remaining $b$-tagged jet(s) and reconstructed $W$'s, taking the combination which gives $m_{bW}$ closest to $m_t$.  If one of the leptonically-decaying $W$ solutions is selected here, the other is discarded. Reject the event if no combination satisfies $|m_{bW}-m_t|\leq 30 {\rm~GeV}$.
\item Reconstruct the charged Higgs candidate from the remaining $W$ and the reconstructed $\hobs$ to determine the discriminating variable $m_{W\hobs}$.
\end{enumerate}

\begin{figure}[t]
\centering
\subfloat[]{%
\label{fig:-a}%
\includegraphics*[angle=0,scale=0.42]{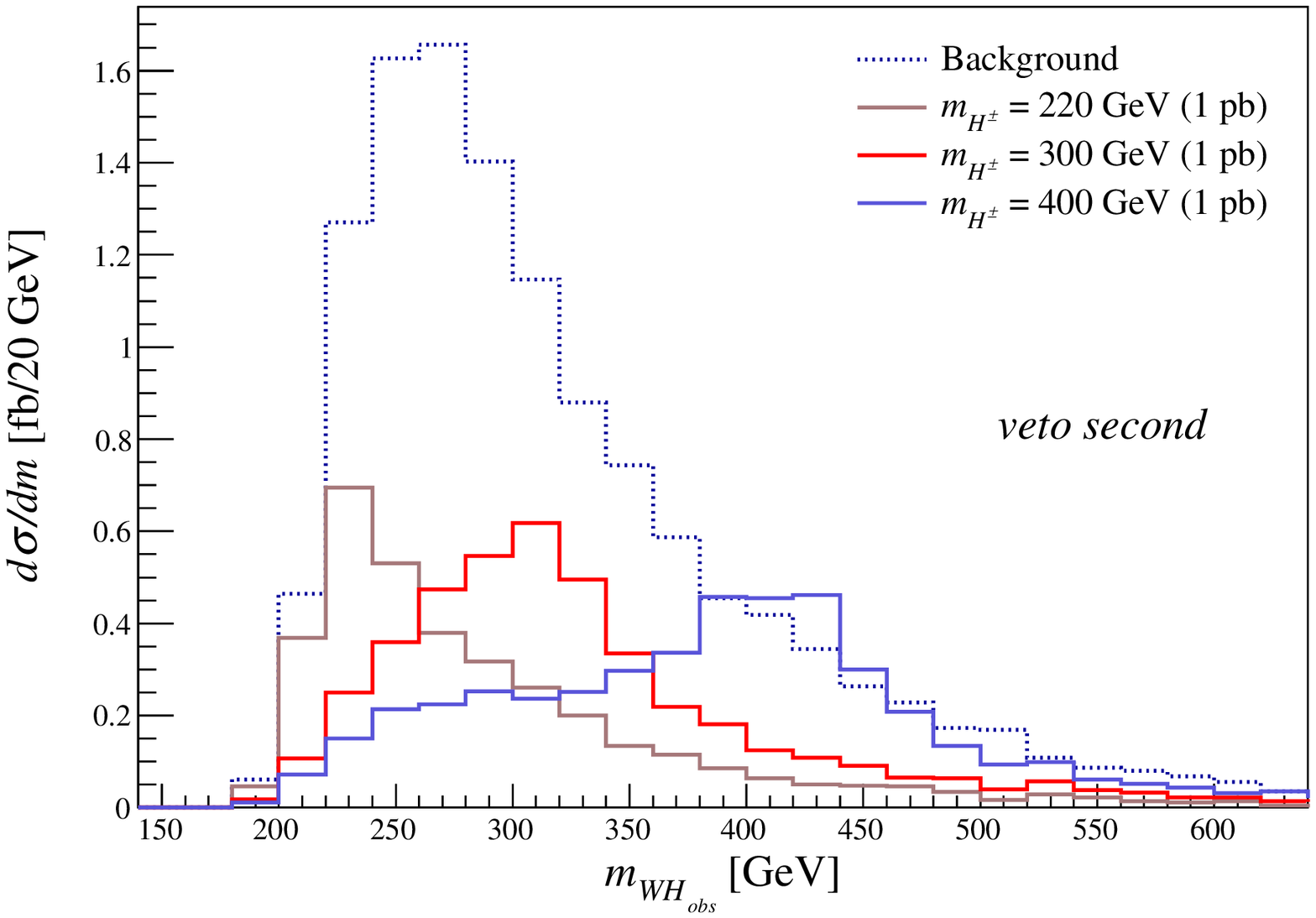}
}%
\subfloat[]{%
\label{fig:-b}%
\includegraphics*[angle=0,scale=0.42]{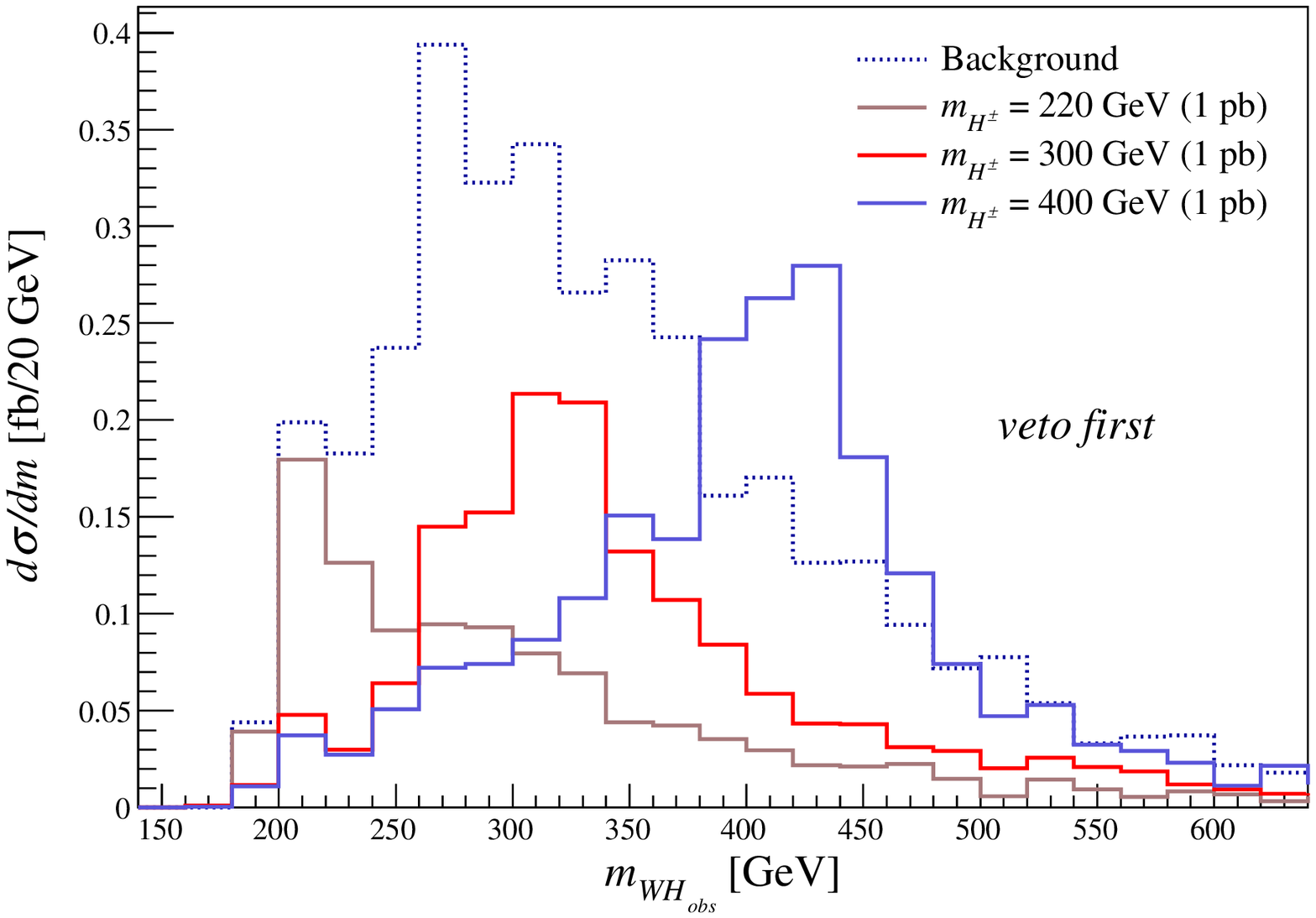}
}%
\caption[]{ Reconstructed $m_{W\hobs}$ for signal and background with
  two different top vetos: (a) first identify an $\hobs \to b\bar{b}$
  candidate, then veto event if two top jets can be reconstructed with
  remaining objects (veto second); (b) using all final state objects,
  veto event if two top jets can be reconstructed (veto first). The
  signal is normalised to
$\sigma(pp\rightarrow tH^\pm)\times
  {\rm BR}(H^\pm\rightarrow W^\pm\hobs) \times {\rm BR}(\hobs\rightarrow
  b\bar{b})=1$\,pb before selection and cuts.}
\label{fig:sigback}
\end{figure}

Because the largest background is by far $t\bar{t}X$, we wish to suppress it as much as possible by identifying events in which a top quark pair can be reconstructed.  The majority of $t\bar{t}X$ events which are able to pass our requirement of providing an SM-like Higgs candidate do so by combining a $b$-jet coming from a top decay with another $b$-tagged jet, so the background will be most reduced if a top veto is applied before the Higgs reconstruction,
\begin{itemize}
\item[] {Veto first:} Using reconstructed $W$'s and all remaining jets, veto event if two top quarks can be reconstructed, both with $|m_{Wj}-m_t|\le 20{\rm~GeV}$.
\end{itemize}
We also wish to avoid unnecessarily cutting signal events.  When a
charged Higgs boson with $m_{H^\pm}\ge m_t$ undergoes the decay
$H^\pm\to W^\pm \hobs\to W^\pm b\bar{b}$, it is kinematically possible
for one of the $b$-jets from the $\hobs$ decay to combine with the $W$
to give an invariant mass close to the top mass.  Indeed, this effect
occurs in large regions of the available phase space for charged Higgs
bosons with masses just above the threshold for $W^\pm \hobs$ decays.  In this case, we wish to identify the $b\bar{b}$ pair from the $\hobs$ decay before applying a top veto,
\begin{itemize}
\item[] {Veto second:} After identifying two $b$-jets which reconstruct $\hobs$, using reconstructed $W$s and all remaining jets, veto event if two top quarks can be reconstructed, both with $|m_{Wj}-m_t|\le 20{\rm~GeV}$.
\end{itemize}
Fig.\,\ref{fig:sigback}(a) and (b) show the signal and background
$m_{W\hobs}$ distributions for $m_{H^\pm}=220,300,400$\gev\ and the two types of top veto. The ``veto first'' scenario clearly reduces the background more effectively, but at the expense of a reduced signal. However, for larger $m_{H^\pm}$, the signal is less likely to fake an additional top, so there is less difference between the two vetoes in the higher mass signal distributions.

\begin{figure}[tbp]
\centering
\includegraphics*[angle=0,scale=0.43]{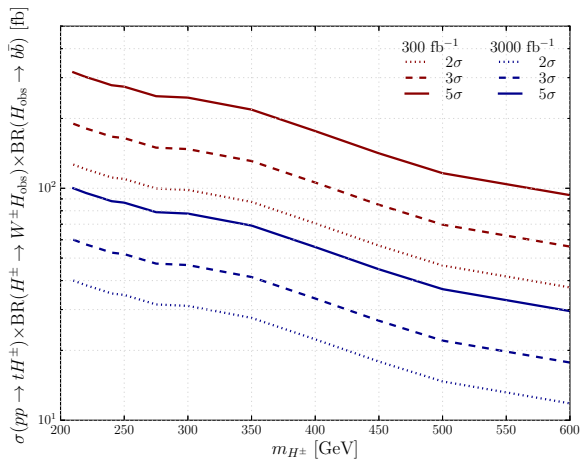}
\caption[]{Sensitivity of the LHC to the signal cross section for exclusion, evidence and
  discovery, based on statistical uncertainties. Contours are thus shown for 
$S/\sqrt{B}=2,3,5$ for an integrated luminosity of $\mathcal{L}=300
$\,fb$^{-1}$ at the next LHC run and at the high luminosity LHC with
$\mathcal{L}=3000$\,fb$^{-1}$, both at $\sqrt{s}=14$\tev. 
}
\label{fig:sensitivity}
\end{figure}

It is also clear from Fig.\,\ref{fig:sigback} that the $H^\pm$
resonance can be reconstructed well enough to further separate it from
the background. For each mass, we select a window in the reconstructed
$m_{W\hobs}$ range which maximises the statistical significance
$S/\sqrt{B}$ of the signal.\footnote{In events where a leptonic $W$
  with two real solutions is used in the reconstruction, the event is
  accepted if either solution gives a $m_{W\hobs}$ within the window.}
We additionally choose the top veto which maximises $S/\sqrt{B}$ for
each mass, and find that ``veto second'' is most effective at lower
masses, $\mhpm\lesssim 350 {\rm~GeV}$, whereas ``veto first'' is
preferable above this mass range.\footnote{As already mentioned, here we consider only statistical uncertainties
(and give
the significance as $S/\sqrt{B}$). A full experimental analysis with all errors
  included might prefer a different mass for the transition between
  vetoes.}  In Fig.\,\ref{fig:sensitivity} we show how this signal and
background translate into sensitivities at the 14 TeV LHC for different
values of the product $\sigma(pp\rightarrow tH^\pm)\times {\rm
BR}(H^\pm\rightarrow W^\pm\hobs)\times {\rm BR}(\hobs\rightarrow
b\bar{b})$, which we henceforth refer to as the signal cross section. 
We see that we can probe 
$\sigma\times {\rm BR}\sim\mathcal{O}$(100\,fb) with an integrated
luminosity of 300\,fb$^{-1}$, but require higher luminosities to see 
$\mathcal{O}$(10\,fb) signals. These sensitivities can be compared to
the model-dependent cross sections and BRs in various scenarios, 
which we discuss in the following section.

\section{\label{results}Results and discussion}

\subsection{MSSM}

\begin{figure}[tbp]
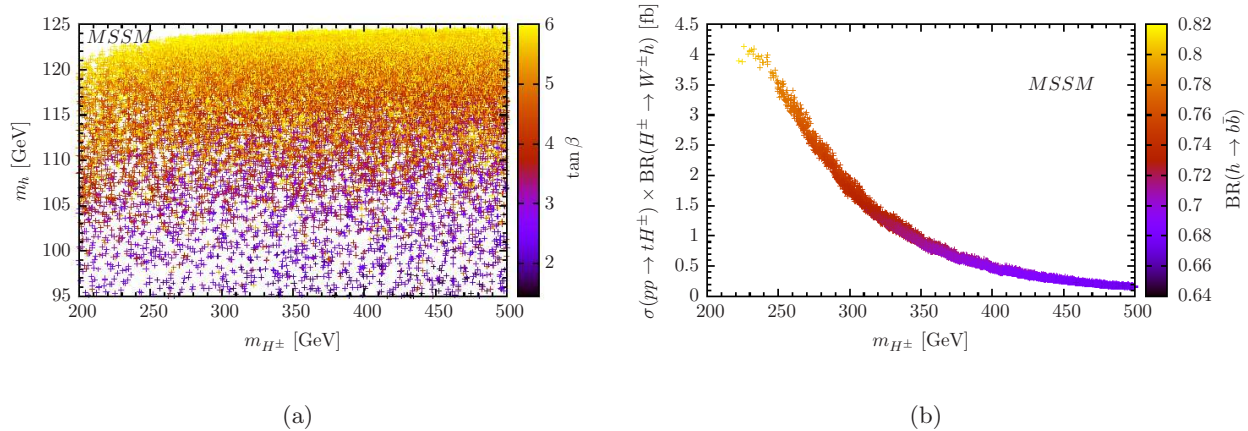

\begin{center}
\vspace*{-0.8cm}
\hspace*{-0.53cm}
\subfloat[]{%
\label{fig:-a}%
\resizebox{0.51\textwidth}{!}{\input{MSSM_mh_mhpm_tanb.tex}}
}%
\hspace{-0.2cm}%
\subfloat[]{%
\label{fig:-b}%
\resizebox{0.51\textwidth}{!}{\input{MSSM_mhpm_XS_mh.tex}}
}%
\caption[]{ (a) $m_h$ as a function of $\mhpm$ in the MSSM, with the
  heat map showing the parameter \tanb. (b) $\sigma(pp\rightarrow t\hpm)\times
  \text{BR}(\hpm\rightarrow W^\pm \hobs)$ as a function of $\mhpm$ in
  the MSSM, with the heat map showing the BR($\hobs\rightarrow b\bar{b}$).}
\label{fig:MSSM}
\end{center}
\end{figure}

In Fig.\,\ref{fig:MSSM}(a) we show the mass of $h$ as a function of
\mhpm\ in the MSSM, with the heat map corresponding to \tanb. The
ranges of the MSSM input parameters scanned to obtain these points are 
shown in Table~\ref{tab:SUSYpara}(a).
One sees in the figure that for the selected \mhpm\ range, \mhsm\ lying
between 122\gev\ -- 128\gev\ can only be obtained for
 $\tanb \gtrsim 6$. As noted earlier, such intermediate values
of \tanb\ bring down not only the $pp \rightarrow t\hpm$ cross section but
also the BR$(H^\pm \rightarrow W^\pm \hobs)$. The product of
these two quantities, only for points in the narrow strip corresponding to
$\mhsm > 122$\gev\ and consequently to highest allowed \tanb\ in
Fig.\,\ref{fig:MSSM}(a), is shown in Fig.\,\ref{fig:MSSM}(b). 
This product hardly exceeds 4\,fb, and that too only for points very close
to the lower limit imposed on \mhsm. The heat map in the figure shows
the BR$(\hobs \rightarrow b\bar{b})$, which grows as the
\hobs\ becomes more and more SM-like due to falling $m_A$, and hence \mhpm,
given the intermediate value of \tanb.

\subsection{NMSSM}

Our initial scans for the NMSSM covered very wide
ranges of the nine input parameters mentioned in \refsec{models}. These scans
revealed only a small region of the NMSSM-specific parameters where
$m_{H_{\rm obs}}$ and \mhpm\ both lied within the desired ranges. 
Two subsequent scans of this narrow region, for the cases with $\hobs = H_1$ and
with $\hobs = H_2$ each, yielded a much larger
density of interesting points. The corresponding parameter
ranges are given in Table~\ref{tab:SUSYpara}(b).

\begin{table}
\begin{center}
\begin{tabular}{cc}
\begin{tabular}{|c|c|}
\hline
MSSM parameter & Range \\
\hline
\mzero\, (GeV) 	& 500 -- 4000	\\
\mhalf\, (GeV)  & 300 -- 2000	\\
\azero\, (GeV)  & $-7000$ -- 7000\\
$\mu$\, (GeV) 	& 100 -- 2000	\\
$m_A$\, (GeV)  	& 100 -- 500	\\
\tanb\ 		& 1 -- 6		\\
\hline
\end{tabular}
&
\hspace*{1cm}\begin{tabular}{|c|c|}
\hline
NMSSM parameter & Range \\
\hline
\mzero\, (GeV) 	& 500 -- 3000	\\
\mhalf\, (GeV)  & 300 -- 2000	\\
\azero\, (GeV)  & $-4000$ -- 4000\\
\tanb\ 		& 1 -- 6 \\
\lam\ 		& 0.45 -- 0.7 \\
\kap\ 		& 0.2 -- 0.5 \\
\mueff\,(GeV) 	& 100 -- 200 \\
\alam\,(GeV)  	& 0 -- 500 \\
\akap\,(GeV)  	& $-500$ -- 0\\
\hline
\end{tabular} \\
(a) & \hspace*{1cm} (b) \\
\end{tabular}
\end{center}
\caption{Ranges of the input parameters scanned for (a) the MSSM and
  (b) the NMSSM.}
\label{tab:SUSYpara}
\end{table}

In Fig.\,\ref{fig:NMSSM-1}(a) we show the BR$(\hpm\rightarrow W^\pm \hobs)$
 as a function of \mhpm\ for the
points obtained in the scan requiring $H_1$ to be the \hobs. 
In Fig.\,\ref{fig:NMSSM-1}(b) the
corresponding points for the case with $\hobs = H_2$ are shown. 
The heat maps in the two figures show the
distribution of the $\sigma(pp\rightarrow t\hpm)$. 
We see in the figures that while the BR$(\hpm\rightarrow W^\pm
\hobs)$ in the $H_1 = \hobs$
 ($H_2 = \hobs$) case can reach up to $\sim 23\%$ ($\sim 28\%$), 
its maximum reachable value drops slowly with decreasing
 $m_{H^\pm}$ and, in fact, for $\mhpm<250$\gev\ it falls below
5\%. This behaviour of the BR$(H^\pm
\rightarrow W^\pm \hobs)$ is thus in conflict with that of the $\sigma(pp\rightarrow
t\hpm)$, which clearly rises with decreasing \mhpm\ and is in fact
maximal for points with the lowest BR$(\hpm\rightarrow W^\pm
\hobs)$ observed. 

In Fig.\,\ref{fig:NMSSM-2}(a) we show the signal cross section for
the case with $\hobs = H_1$. The points in green are the ones fulfilling only
 the $b$-physics constraints and we note for these points that, as 
a result of the tension between the BR$(H^\pm
\rightarrow W^\pm \hobs)$ and the $\sigma(pp\rightarrow
t\hpm)$, the total cross section barely exceeds 10\,fb. The points in
red and blue in the figure are the ones for which
$R^{\gamma\gamma/ZZ}$ are consistent with the CMS and ATLAS ranges of 
$\mu^{\gamma\gamma/ZZ}$, respectively. Evidently, imposing these constraints
further reduces the maximum signal cross section obtainable to below
5\,fb. For the case with $\hobs = H_2$ the signal cross section, shown in
Fig.\,\ref{fig:NMSSM-2}(b), can reach slightly higher to around
20\,pb, for the green points. This is owing to the somewhat larger BR$(H^\pm
\rightarrow W^\pm \hobs)$ obtainable for low \mhpm\ in this case
compared to the $\hobs = H_1$ case. However, again the overall signal cross
section is highly diminished for points observing the ATLAS or CMS
signal rate constraints. Also shown in the
Figs.\,\ref{fig:NMSSM-2}(a) and (b) are the 2$\sigma$ (exclusion),
3$\sigma$ (evidence) and 5$\sigma$ (discovery) sensitivity curves for
3000\,fb$^{-1} $ accumulated luminosity at the LHC 14\,TeV run. All
the good points from the scans lie well below the lowest (2$\sigma$) curve,
implying that none of them has a signal cross section large enough to be testable
even at such a high luminosity.  

\begin{figure}[ht!]
\begin{center}
\vspace*{-1cm}
\hspace*{-0.53cm}
\subfloat[]{%
\label{fig:-a}%
\resizebox{0.51\textwidth}{!}{\input{./h1NMSSM_mhpm_BRhpmh1w_sigma.tex}}
}%
\subfloat[]{%
\label{fig:-b}%
\resizebox{0.51\textwidth}{!}{\input{./h2NMSSM_mhpm_BRhpmh2w_sigma.tex}}
}%
\caption[]{ BR$(\hpm\rightarrow W^\pm \hobs)$ as a function of \mhpm\ in
  the NMSSM when (a) $\hobs =  H_1$ and (b) $\hobs = H_2$, with the
  heat map showing the $\sigma(pp\rightarrow t\hpm)$.}
\label{fig:NMSSM-1}
\end{center}
\end{figure}

\begin{figure}[ht!]
\begin{center}
\hspace*{-0.5cm}
\subfloat[]{%
\resizebox{0.515\textwidth}{!}{\input{./h1NMSSM_mhpm_XS.tex}}
}%
\subfloat[]{%
\resizebox{0.515\textwidth}{!}{\input{./h2NMSSM_mhpm_XS.tex}}
}%
\caption[]{Signal cross section as a function of \mhpm\ in the NMSSM when (a) $\hobs =  H_1$ and (b) $\hobs = H_2$. See text for details.}
\label{fig:NMSSM-2}
\end{center}
\end{figure}

\subsection{2HDM Types I and II}

The scanned ranges of the parameters in these two models are shown in
Table~\ref{tab:2hdmpara}. Note that in the 2HDM-II, 
$\mhpm \lesssim 320$\gev\ is excluded for all values of \tanb\ by the 
constraint on \brbxsgamma, while $\tanb \lesssim 1.5$ is ruled
out for \mhpm\ up to 500\gev\ or so by the $\Delta M_{B_d}$ constraint, according 
to\cite{Mahmoudi:2009zx}. We therefore reduced the input range of \mhpm\
instead of imposing these constraints during the scans
for this model. The BR$(\hpm\rightarrow W^\pm \hobs)$ for 2HDM-I
with the $\hobs = h$ case, shown in Fig.\,\ref{fig:T12HDM-1}(a), can
be as high as $\sim 95\%$ for a fairly large number of points. 
Moreover, compared to the NMSSM, while the maximum
$\sigma(pp\rightarrow t\hpm)$ 
reachable is much lower here, the BR$(\hpm\rightarrow W^\pm \hobs)$
grows much more sharply with increasing \mhpm. As a result, there are 
plenty of low \mhpm\
points where both the BR$(\hpm\rightarrow W^\pm \hobs)$ as well as the
$\sigma(pp\rightarrow t\hpm)$, shown by the heat map, can be
significant. In
Fig.\,\ref{fig:T12HDM-1}(b) are shown the corresponding quantities for
the $\hobs = H$ case in the 2HDM-I. In this case a very large
BR$(\hpm\rightarrow W^\pm \hobs)$ is obtainable for a comparatively
much smaller number of points and it mostly stays below
40\%.

\begin{table}
\begin{center}
\begin{tabular}{|c|c|c|c|c|}
\hline
\multirow{2}{*}{Parameter} & \multicolumn{2}{c|}{2HDM-I} & \multicolumn{2}{c|}{2HDM-II} \\ 
                 & $\hobs = h$ & $\hobs = H$ & $\hobs = h$ & $\hobs = H$ \\
\hline
$m_h$\,(GeV) 		& 123 -- 127 & 80 -- 115  & 123 -- 127 & 80 -- 115  \\
$m_H$\,(GeV) 		& 135 -- 500 & 123 -- 127 & 135 -- 500 & 123 -- 127 \\
$\mhpm=m_A$\,(GeV) 	& \multicolumn{2}{c|}{135 -- 500} & \multicolumn{2}{c|}{320 -- 500}  \\
\tanb			& \multicolumn{4}{c|}{1.5 -- 6}  \\
$|\sin (\beta - \alpha)|$ & \multicolumn{4}{c|}{0 -- 1}  \\
$m_{12}^2$\,(GeV$^2$) & \multicolumn{4}{c|}{0 -- $m_A^2 \cos \beta \sin \beta$ }  \\
\hline
\end{tabular}
\end{center}
\caption{Ranges of the input parameters scanned for the 2HDM Types I and II.}
\label{tab:2hdmpara}
\end{table}

\begin{figure}[ht]
\begin{center}
\vspace*{-0.5cm}
\hspace*{-0.53cm}
\subfloat[]{%
\label{fig:-a}%
\resizebox{0.51\textwidth}{!}{\input{./h1t12hdm_mhpm_BRhpmhw_sigma.tex}}
}%
\subfloat[]{%
\label{fig:-a}%
\resizebox{0.51\textwidth}{!}{\input{./h2t12hdm_mhpm_BRhpmhw_sigma.tex}}
}%
\caption[]{ BR$(\hpm\rightarrow W^\pm \hobs)$ as a function of \mhpm\ in
  the 2HDM-I when (a) $\hobs =  h$ and (b) $\hobs = H$, with the
  heat map showing the $\sigma(pp\rightarrow t\hpm)$.}
\label{fig:T12HDM-1}
\end{center}
\end{figure}

\begin{figure}[ht]
\begin{center}
\hspace*{-0.5cm}
\subfloat[]{%
\label{fig:-a}%
\resizebox{0.515\textwidth}{!}{\input{./h1t12hdm_mhpm_XS.tex}}
}%
\subfloat[]{%
\label{fig:-a}%
\resizebox{0.515\textwidth}{!}{\input{./h2t12hdm_mhpm_XS.tex}}
}%
\caption[]{Signal cross section as a function of \mhpm\ in the
  2HDM-I when (a) $\hobs =  h$ and (b) $\hobs = H$. See text for details.}
\label{fig:T12HDM-2}
\end{center}
\end{figure}

\begin{figure}[ht]
\begin{center}
\vspace*{-1cm}
\hspace*{-0.53cm}
\subfloat[]{%
\label{fig:-a}%
\resizebox{0.51\textwidth}{!}{\input{./h1t22hdm_mhpm_BRhpmhw_sigma.tex}}
}%
\subfloat[]{%
\label{fig:-a}%
\resizebox{0.51\textwidth}{!}{\input{./h2t22hdm_mhpm_BRhpmhw_sigma.tex}}
}%
\caption[]{ BR$(H^-\rightarrow W^-\hobs)$ as a function of $m_{H^\pm}$ in
  the 2HDM-II when (a) $\hobs =  h$ and (b) $\hobs = H$, with the heat map
  showing the $\sigma(pp\rightarrow t\hpm)$.}
\label{fig:T22HDM-1}
\end{center}
\end{figure}

\begin{figure}[ht]
\begin{center}
\hspace*{-0.5cm}
\subfloat[]{%
\label{fig:-a}%
\resizebox{0.515\textwidth}{!}{\input{./h1t22hdm_mhpm_XS.tex}}
}%
\subfloat[]{%
\label{fig:-a}%
\resizebox{0.515\textwidth}{!}{\input{./h2t22hdm_mhpm_XS.tex}}
}%
\caption[]{Signal cross section as a function of \mhpm\ in the
  2HDM-II when (a) $\hobs =  h$ and (b) $\hobs = H$. See text for details.}
\label{fig:T22HDM-2}
\end{center}
\end{figure}

In Fig.\,\ref{fig:T12HDM-2}(a) we show the signal cross section for
the $\hobs = h$ case in the 2HDM-I as a function of \mhpm. The colour
convention for the points in all the figures showing the signal cross
section henceforth is the same as in Fig.\,\ref{fig:NMSSM-2}.  We note that,
owing to the much larger BR$(\hpm\rightarrow W^\pm \hobs)$ generally 
obtainable in this model compared to the NMSSM, the total cross
section can reach as high as about 100\,fb. A small portion of the
green points with $\mhpm > 400$\gev\ lies above the 2$\sigma$
sensitivity curve corresponding to $\mathcal{L} = 300$\,fb$^{-1}$ and
should thus be reachable at the LHC. The picture, however, becomes grim
when the LHC signal rate constraints are imposed. Points consistent
with the CMS constraints have a maximum possible cross section of 
around 20\,fb, while none of the points obtained in the scans
are able to satisfy the ATLAS constraints. 
 
Turning to the 2HDM-II, for the $\hobs = h$ case one sees in 
Fig.\,\ref{fig:T22HDM-1}(a) that in this model both 
the BR$(\hpm\rightarrow W^\pm \hobs)$ and 
the $\sigma(pp\rightarrow t\hpm)$ show a similar behaviour as noted in the
2HDM-I above, being significantly large simultaneously for a number of
points with \mhpm\ up to $\sim 400$\gev. The maximum obtainable values
 of both these quantities are also similar to those in the 2HDM-I. 
In the $\hobs = H$ case 
the BR$(\hpm\rightarrow W^\pm \hobs)$ struggles to reach high values
generally and in fact stays close to 0 for a vast majority of the points,
as seen in Fig.\,\ref{fig:T22HDM-1}(b). In 
Figs.\,\ref{fig:T22HDM-2}(a) and (b) we show the signal cross sections
for the $\hobs = h$ and $\hobs = H$ cases, respectively, in the
2HDM-II. In the former case, not only do a large number of points
observing only the $b$-physics constraints lie above the 5$\sigma$
sensitivity curve for $\mathcal{L} = 3000$\,fb$^{-1}$, but also some of
the points consistent with the CMS constraints can have a signal cross
section in excess of 30\,fb and should thus be accessible at the LHC. In the $\hobs =
H$ case, however, the maximum reachable cross section for points
consistent with the CMS and ATLAS signal rate constraints barely
exceeds 10\,fb and 1.5\,fb, respectively, only when \mhpm\ is below 350\gev\
or so.

\subsection{A2HDM}

The scanned ranges of the A2HDM parameters are given in
Table~\ref{tab:a2hdmpara} and have been adopted from\cite{Enberg:2013jba}.
In Fig.\,\ref{fig:Al2HDM-1}(a) we show the BR$(\hpm\rightarrow W^\pm
\hobs)$ for the $\hobs = h$ case, which can reach unity over the
entire desired mass range of \hpm. Also, the $\sigma(pp\rightarrow t\hpm)$,
illustrated by the heat map in the figure, can reach the pb level, 
but it is maximal only for points for which the 
BR$(\hpm\rightarrow W^\pm \hobs)$ is relatively small, $\lesssim 40\%$.
On the other hand, Fig.\,\ref{fig:Al2HDM-1}(b) shows that in the
$\hobs = H$ case the BR$(\hpm\rightarrow W^\pm \hobs)$
mostly stays below $\sim 35$\%.

\begin{figure}[ht!]
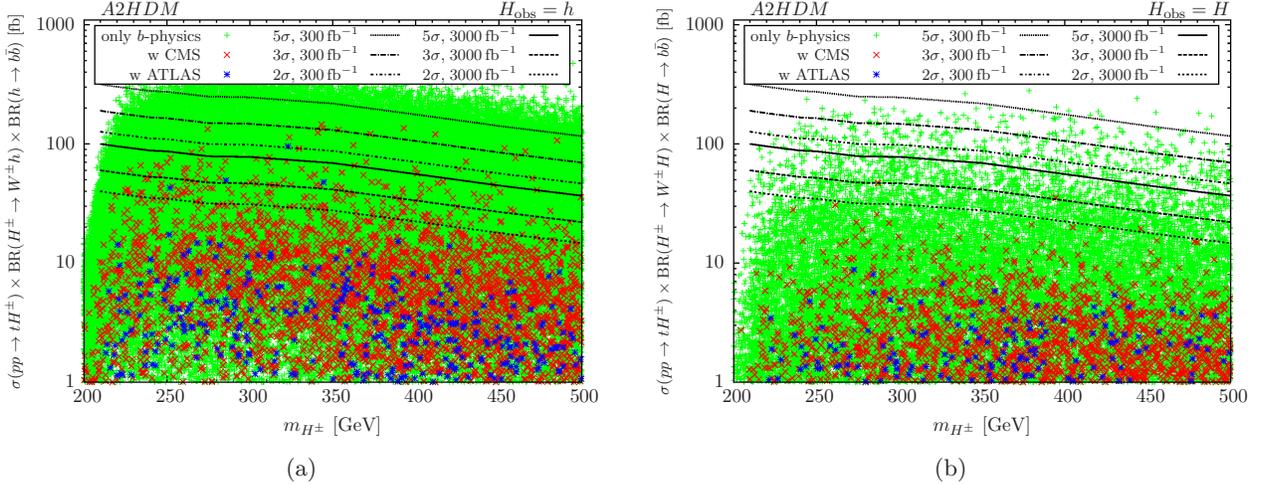

\begin{center}
\hspace*{-0.5cm}
\subfloat[]{%
\label{fig:-a}%
\resizebox{0.515\textwidth}{!}{\input{./h1al2hdm_mhpm_XS.tex}}
}%
\subfloat[]{%
\label{fig:-a}%
\resizebox{0.515\textwidth}{!}{\input{./h2al2hdm_mhpm_XS.tex}}
}%
\caption[]{Signal cross section as a function of \mhpm\ in the
  A2HDM when (a) $\hobs =  h$ and (b) $\hobs = H$. See text for details.}
\label{fig:Al2HDM-2}
\end{center}
\end{figure}

In Fig.\,\ref{fig:Al2HDM-2}(a) the signal cross section for
the $\hobs = h$ case is shown. This cross section can reach much higher, $\sim
700$\,fb, than in the ordinary 2HDMs, when the constraints
from the LHC Higgs boson searches are not imposed. Points with such a
high cross section lie above even the 5$\sigma$ sensitivity curve for 
 the LHC with $\mathcal{L} = 300$\,fb$^{-1}$.
This implies that the \hpm\ in this model could be discoverable at the standard
luminosity LHC over almost the entire mass range
analysed for this channel. However, as in the other models above, points
satisfying the LHC constraints have a
much smaller signal cross section generally. Still, unlike in any of the
other models considered here, a small number of points consistent with
the CMS constraints lies  above the 5$\sigma$ sensitivity curve for 
$\mathcal{L} = 3000$\,fb$^{-1}$ and could thus be visible at the high
luminosity LHC. The same is not true though for the $\hobs = H$
case, seen in Fig.\,\ref{fig:Al2HDM-2}(b), where only a couple of
points consistent with the CMS constraints appear to be testable 
at the high luminosity LHC.  \\

\begin{table}
\begin{center}
\begin{tabular}{|c|c|c|}
\hline
Parameter  & $H_\text{obs} = h $&  $H_\text{obs} = h $\\ \hline
$m_h$ (GeV) & 123 -- 127 & 80 -- 115 \\
$m_H$ (GeV) & 135 -- 300 & 123 -- 127 \\
$m_{H^\pm} = m_A$ (GeV) & \multicolumn{2}{c|}{200 -- 500} \\
$| \sin \alpha |$ &  \multicolumn{2}{c|}{0 -- 1} \\
$ \lambda_2 $ &  \multicolumn{2}{c|}{ 0 -- $4\pi$} \\
$ \lambda_3 $ &  \multicolumn{2}{c|}{$-\sqrt{\lambda_1 \, \lambda_2 \, }$ -- $4\pi$} \\
$ |\lambda_7| $ &  \multicolumn{2}{c|}{0 -- $4\pi$} \\
$ |\beta^{U,D,L}|$ & \multicolumn{2}{c|}{0 -- 1.57} \\
\hline
\end{tabular}
\end{center}
\caption{Ranges of the input parameters scanned for the A2HDM.}
\label{tab:a2hdmpara}
\end{table}

\begin{figure}[ht!]
\begin{center}
\vspace*{-1cm}
\hspace*{-0.53cm}
\subfloat[]{%
\label{fig:-a}%
\resizebox{0.51\textwidth}{!}{\input{./h1al2hdm_mhpm_BRhpmhw_sigma.tex}}
}%
\subfloat[]{%
\label{fig:-a}%
\resizebox{0.51\textwidth}{!}{\input{./h2al2hdm_mhpm_BRhpmhw_sigma.tex}}
}%
\caption[]{ BR$(H^-\rightarrow W^-\hobs)$ as a function of $m_{H^\pm}$ in
 the A2HDM when (a) $\hobs =  h$ and (b) $\hobs = H$, with the heat map
  showing the $\sigma(pp\rightarrow t\hpm)$.}
\label{fig:Al2HDM-1}
\end{center}
\end{figure}

\section{\label{concl} Conclusions}

In this article we have analysed the detectability of $\hpm$ 
in the $W\hobs$ decay mode in some minimal extensions of the SM, 
at the upcoming Run 2 of the LHC with $\sqrt{s}=14$\tev. 
We have discussed some important features of the models of our interest,
in particular the coupling parameters governing the production of
$\hpm$ in $pp$ collisions as well as the $\hpm \rightarrow W\hobs$ 
decay process.  We have performed dedicated scans of the parameter
spaces of these models to search for their regions where a \hpm\ with a mass lying in
the 200\gev\ -- 500\gev\ range can be obtained and its production
cross section can be maximised. These scans were subject to the most
relevant constraints from $b$-physics, from the LHC Higgs boson
searches and, in the case of SUSY models, from relic density
measurements. Moreover, in the NMSSM as well as in the 2HDMs we
considered both the possibilities of the observed Higgs boson being
the lightest or the next-to-lightest CP-even scalar of the model.

We then reconstructed the signal and the background in the
$bbb(b)jj\ell\nu_\ell$ final state and, through a dedicated
detector-level analysis, estimated the signal significance for various
accumulated luminosities at the LHC. We found that, through a judicious choice of selection criteria, including a veto on $t\bar{t}$ events and the requirement of a reconstructed $125\gev$ Higgs boson from a pair of $b$-tagged jets, we were able to significantly reduce the backgrounds.  The semi-leptonic channel provides enough kinematic information to reconstruct the $\mhpm$ peak and identify signals with $\sigma(pp\rightarrow tH^\pm)\times {\rm BR}(H^\pm\rightarrow W^\pm\hobs)\times {\rm BR}(\hobs\rightarrow b\bar{b})\sim\mathcal{O}$(100\,fb) with an integrated
luminosity of 300\,fb$^{-1}$, with even better sensitivity at high luminosities.

We have concluded that in the SUSY models studied here, the $\hpm
\rightarrow W\hobs$ decay channel does not
carry as much promise for the identification of a \hpm\
as has been envisaged in some earlier studies. This is due to the fact
that the $pp\rightarrow \hpm$ production process 
and the subsequent $\hpm\rightarrow W\hobs$ decay process generally 
show contrasting 
dependence on the various parameters involved. 
The situation looks a bit better in the $Z_2$-symmetric 2HDMs, as long
as the constraints from the LHC measurements of the Higgs boson signal
rates are ignored. Imposing these constraints leaves an insignificant
number of  points in the 2HDM-II visible at only the high luminosity ($\sim
3000$\,fb$^{-1}$) LHC, implying that the Higgs boson assumed to be the
one observed at the LHC in these scenarios deviates substantially 
from SM-like properties. In the case of the A2HDM, a fairly large portion
of the parameter space could in general be tested even at the standard 
luminosity ($\sim 300$\,fb$^{-1}$) LHC. However, again if the
measurements of the observed Higgs boson signal rates do not
fluctuate much from the current ones, only a few parameter space
points lie within the reach of the LHC at this luminosity. 
   

\begin{center}
  \textbf{ACKNOWLEDGEMENTS}
\end{center}

\noindent 
This work was in part funded by the Swedish Research Council under contracts 2007-4071 and 621-2011-5107.
The work of S.~Moretti has been funded in part through the NExT Institute. 
The computational work was in part carried out on
resources provided by the Swedish National Infrastructure
for Computing (SNIC) at Uppsala Multidisciplinary Center
for Advanced Computational Science (UPPMAX) under Projects p2013257 and SNIC 2014/1-5.


\bibliographystyle{utphysmcite}	

\bibliography{Hpm_WH_refs}
\end{document}